%% file: static.tex
\documentstyle[11pt,amsfonts,fullpage]{amsart}
\def\endproof{\qed\smallskip}
\def\blacksquare{\hbox to .60em{\vrule width .60em height .60em}}

\newtheorem{theorem}{Theorem}[section] 
 
\newtheorem{remark}[theorem]{Remark}

\newcommand{\Roof}[2]{#2{#1}}

\newcommand{\bbgin}{\begin}

\begin{document}

\title[]{On the Structure of Solutions to the Static 
Vacuum Einstein Equations}

\author[]{Michael T. Anderson}

\thanks{Partially supported by NSF Grants DMS 9505744 and 9802722}

\maketitle
\setcounter{section}{-1}
\section{Introduction}

\setcounter{equation}{0}

 The static vacuum Einstein equations are the equations
\begin{equation} \label{e0.1}
ur = D^{2}u, 
\end{equation}
$$\Delta u = 0, $$
on a Riemannian 3-manifold $(M, g)$, with $u$ a positive function on 
$M$. Here $r$ denotes the Ricci curvature, $D^{2}$ the Hessian, and 
$\Delta  = trD^{2}$ the Laplacian on $(M, g)$. Solutions of these 
equations define a Ricci-flat 4-manifold $N$, of the form $N = 
M\times_{u}S^{1}$ or $N = M\times_{u}{\Bbb R}$, with Riemannian or 
Lorentzian metric of the form
\begin{equation} \label{e0.2}
g_{N} = g_{M} \pm  u^{2}dt^{2}. 
\end{equation}

 These equations are the simplest equations for Ricci-flat 4-manifolds. 
They have been extensively studied in the physics literature on 
classical relativity, where the solutions represent space-times outside 
regions of matter which are translation and reflection invariant in the 
time direction $t$. However, with the exception of some notable 
instances, (c.f. Theorem 0.1 below), many of the global properties of 
solutions have not been rigorously examined, either from mathematical 
or physical points of view, c.f. [Br] for example. 

 This paper is also motivated by the fact that solutions of the static 
vacuum equations arise in the study of degenerations of Yamabe metrics 
(or metrics of constant scalar curvature) on 3-manifolds, c.f. [A1]. 
Because of this and other related applications of these equations to 
the geometry of 3-manifolds, we are interested in general mathematical 
aspects of the equations and their solutions which might not be 
physically relevant; for example, we allow solutions with negative mass.

 In this paper, we will be mostly concerned with the geometry of the 
3-manifold solutions $(M, g, u)$ of (0.1), (i.e. the space-like 
hypersurfaces), and not with the 4-manifold metric. Thus, the choice of 
Riemannian or Lorentzian geometry on $N$ in (0.2) will play no role. 
This considerably simplifies the discussion of singularities and 
boundary structure, but still allows for a large variety of behaviors; 
c.f. [ES] for a survey on singularities of space-times.

 Obviously, there are no non-flat solutions to (0.1) on closed 
manifolds, and so it will be assumed that $M$ is an open, connected 
oriented 3-manifold. Let $\Roof{M}{\bar}$ be the metric (or Cauchy) 
completion of $M$ and $\partial M$ the metric boundary, so that 
$\Roof{M}{\bar} = M\cup\partial M$ is complete as a metric space.

 In order to avoid trivial ambiguities, we will only consider maximal 
solutions of the equations (0.1). For example any domain $\Omega $ in 
${\Bbb R}^{3}$ with the flat metric, and $u$ a positive constant, 
satisfies (0.1). In this case, the metric boundary $\partial\Omega $ is 
artificial, and has no intrinsic relation with the geometry of the 
solution. The solution obviously extends to a larger domain, i.e. 
${\Bbb R}^{3}.$ Thus, we only consider maximal solutions $(M, g, u)$, 
in the sense that $(M, g, u)$ does not extend to a larger domain $(M' , 
g' ) \supset  (M, g)$ with $u > $ 0 on $M' .$ It follows that at the 
metric boundary $\partial M$ of $M$, either the metric or $u$ 
degenerates in some way or $u$ approaches 0 in some way, (or a 
combination of such).

\medskip

 A classical result of Lichnerowicz [L1, p.137] implies that if the 
metric $(M, g)$ satisfying (0.1) is complete, (i.e. $\partial M = 
\emptyset  ),$ and $u \rightarrow $ 1 at infinity, then $u \equiv $ 1 
and $M$ is flat, i.e. ${\Bbb R}^{3}$ or a quotient of ${\Bbb R}^{3}.$ 
More generally, it is proved in [A1, Thm.3.2] that if $(M, g)$ is a 
complete solution to (0.1), (hence $u > $ 0 everywhere), then $(M, g)$ 
is flat and $u$ is constant, i.e. the assumption on the asymptotic 
behavior of $u$ is not necessary, c.f. also Theorem 1.1 below. 

 Thus, there are no complete non-trivial solutions to (0.1) and hence 
$\partial M$ must be non-empty. The set formally given by 
$$\Sigma  = \{u = 0\} \subset  \bar M,$$ 
is called the {\it horizon}. It is closely related to the notion of 
event horizon in general relativity. More precisely, $\Sigma$ may be 
defined as the set of limit points of Cauchy sequences on $(M, g)$ on 
which $u$ converges to $0$. Although $\partial M \neq \emptyset$, it is 
possible that $\Sigma = \emptyset$. However, most solutions of physical 
interest do have $\Sigma \neq \emptyset$. In the framework of classical 
relativity, the non-triviality of a static vacuum solution, i.e. the 
non-vanishing of its curvature, is due to the presence of matter or 
field sources at $\partial M,$ or 'inside` the horizon $\Sigma $ in 
case $(M, g)$ extends as a vacuum solution past $\Sigma .$

 It is natural to consider the situation where $(M, g)$ is not complete 
and for which the metric boundary $\partial M$ of $M$ coincides with 
the horizon $\Sigma$. More precisely, we will say that $(M, g)$ is {\it 
 complete away from the horizon} $\Sigma$ if for any sequence $p_{i} 
\rightarrow  p \in\partial M$ in the metric topology on $\Roof{M}{\bar} 
= M\cup\partial M$ one has $u(p_{i}) \rightarrow $ 0. Conversely if 
$\{p_{i}\}$ is a bounded sequence in $M$ with $u(p_{i}) \rightarrow $ 
0, then the definition of $(M, g, u)$ implies that a subsequence of 
$\{p_{i}\}$ converges to a point $p\in\partial M.$ Thus, $\partial M = 
\Sigma $ is given by the Hausdorff limit of the $\varepsilon$ -levels 
$L^{\varepsilon}$ of $u$ as $\varepsilon  \rightarrow $ 0. 

 While most solutions $(M, g)$ of physical interest are complete away 
from $\Sigma ,$ there are many solutions for which this is not the 
case, c.f. \S 2 for further discussion. In such examples, the curvature 
typically blows up within a finite distance to $\Sigma .$ Among the 
solutions which are complete away from $\Sigma ,$ most all are singular 
at $\Sigma ,$ again in the sense that the curvature of the metric $g$ 
blows up on approach to $\Sigma .$ This is closely related to the fact 
that the equations (0.1) are formally degenerate at $\Sigma .$ Thus, in 
general, even when $M$ is complete away from $\Sigma ,$ the metric 
completion $M\cup\Sigma $ need not be a smooth manifold with boundary.

 We will say that $(M, g, u)$ {\it  extends smoothly to}  $\Sigma ,$ if 
(i): the set $\Sigma $ is a smooth surface and the partial completion 
$M'  = M\cup\Sigma  \subset  \Roof{M}{\bar}$ is a smooth manifold with 
boundary, (ii): the metric $g$ extends smoothly to a smooth Riemannian 
metric on $M' ,$ and (iii): the potential $u$ extends smoothly to 
$\Sigma $ with $\Sigma  = \{u=0\};$ smoothness here means at least 
$C^{2}.$ Note that it might be possible that $\Sigma $ has infinitely 
many components, or non-compact components of infinite topological 
type. In any case, one immediate strong consequence of (0.1) is that if 
$g$ extends smoothly to $\Sigma ,$ then at $\Sigma $ one has
\begin{equation} \label{e0.3}
D^{2}u|_{\Sigma} = 0. 
\end{equation}
Thus, $\Sigma $ is a totally geodesic surface in $\Roof{M}{\bar}$ and 
$|\nabla u|$ is a non-zero constant on each component of $\Sigma ,$ 
c.f. Remark 1.5 below. Observe that if $(M, g)$ is smooth up to $\Sigma 
,$ and complete away from $\Sigma ,$ then the isometric double of $M$ 
across $\Sigma $ is a smooth complete Riemannian manifold. The harmonic 
function $u$ extends smoothly across $\Sigma $ as harmonic function, 
odd w.r.t. reflection in $\Sigma .$

\medskip

 By far the most significant solution of the static vacuum equations is 
the Schwarzschild metric, of mass $m$, given on the space-like 
hypersurface $M$ by
\begin{equation} \label{e0.4}
g_{S} = (1 - \frac{2m}{r})^{-1}dr^{2} + r^{2}ds^{2}_{S^{2}} , u = (1 - 
\frac{2m}{r})^{1/2}, r >  2m. 
\end{equation}

 This metric models the vacuum exterior region of an isolated static 
star or black hole. It is a spherically symmetric metric on $M$ = 
$(2m,\infty )\times S^{2}$ and has $\Sigma $ given by a (totally 
geodesic) symmetric $S^{2},$ of radius $2m$. The mass $m$ is usually 
assumed to be positive, but we will not make this assumption here. 
Thus, we allow $m \leq $ 0. Of course if $m =$ 0, then $g_{S}$ is just 
the flat metric with $u =$ 1. If $m < $ 0, (0.4) is understood to be 
defined for $r > $ 0. 

 The Schwarzschild metric is asymptotically flat in the sense that 
there is a compact set $K$ in $M$ such that $M\setminus K$ is 
diffeomorphic to ${\Bbb R}^{3}\setminus B(R),$ and in a suitable chart 
on $M\setminus K$, the metric approaches the Euclidean metric at a rate 
of $1/r,$ i.e.
\begin{equation} \label{e0.5}
g_{ij} = (1 + \frac{2m}{r})\delta_{ij} + O(1/r^{2}), 
\end{equation}
with curvature decay of order $1/r^{3}, r = |x|,$ and with $m\in{\Bbb 
R} .$
The function $u$ (up to a multiplicative constant) has the asymptotic 
form
$u =$ 1 $-  \frac{m}{r} + O(r^{-2})$ with $|\nabla u| = O(1/r^{2}).$ A 
triple
$(M,g,u)$ satisfying these conditions is called {\it asymptotically 
flat}. 

 We note the following remarkable characterization of the Schwarzschild 
metric.

\bbgin{theorem} \label{t 0.1.}
  (Black-hole uniqueness),{\rm [I1],[Ro],[BM]}

 Let $(M, g, u)$ be a solution of the static vacuum Einstein equations, 
which is smooth up to $\Sigma $ and complete away from $\Sigma .$ If 
$\Sigma $ is a compact, (possibly disconnected) surface and $(M, g, u)$ 
is asymptotically flat, then $(M, g, u)$ is the Schwarzschild metric, 
with $m > 0$.
\end{theorem}

 The hypothesis that the space-like hypersurface $(M, g, u)$ is 
asymptotically flat is very common in physics. Namely, in modeling a 
static space-time outside an isolated, i.e. compact, field or matter 
source, it is natural to assume that in regions far away from the 
source the geometry of space approximates that of ${\Bbb R}^{3},$ i.e. 
empty space. 

 Nevertheless, mathematically the asymptotically flat assumption is 
quite strong in that it severely restricts both the topology and 
geometry of $(M, g)$ outside a large compact set. Further, the physical 
reasoning above presupposes that there are no complete non-flat 
solutions to the vacuum equations (0.1), i.e. that a static 
gravitational field is non-empty {\it solely} due the presence of 
matter somewhere.

\medskip
\bbgin{remark} \label {Remark 0.2.}
{\rm This latter issue in fact led Einstein to hypothesize that space 
$M$ is compact, in order to avoid dealing with 'artificial` boundary 
value problems at infinity, c.f. [E, p.98ff.]. This issue, closely 
related to Mach's Principle, is discussed in some detail in work of 
Lichnerowicz, (Propositions A and $B$ of [L1, \S 31] and [L2, Ch.II]); 
c.f. also [MTW, \S 21.12,],[Ri, \S 9.12] for instance.}
\end{remark}

\medskip

As remarked above, there are in fact no complete non-trivial static 
vacuum solutions, so that the asymptotically flat assumption in Theorem 
0.1 may not be unreasonable, c.f. however [El], [G]. In fact, the main 
result of this paper is that this hypothesis is not necessary in most 
circumstances; it follows from much weaker assumptions.

 To explain this, we first need to consider a weakening of the 
condition that $\partial M$ is compact. Let
\begin{equation} \label{e0.6}
t(x) = dist(x, \partial M). 
\end{equation}
Define $\partial M$ to be {\it pseudo-compact}  if there is a tubular 
neighborhood $U$ of $\partial M$ whose boundary $\partial U$ in $M$, 
$\partial U\cap M$, is compact, i.e. $\{t(x) = s_{o}\}$ is compact, for 
some $s_{o} > $ 0, (and hence all $0 < s_{o} < \infty$). As will be 
seen in \S 2, there are numerous examples of static vacuum solutions 
with $\partial M$ pseudo-compact but not compact.

 Let $E$ be an end, i.e. an unbounded component, of 
$\Roof{M}{\bar}\setminus U.$ The mass of $E$ may be defined by
\begin{equation} \label{e0.7}
m_{E} = lim_{s \rightarrow \infty}m_{E}(s) = lim_{s \rightarrow 
\infty}{\frac{1}{4 \pi}}\int_{S_{E}(s)}<\nabla logu, \nabla t> dA, 
\end{equation}
where $S_{E}(s) = t^{-1}(s)\cap E$. Since $u$ is
harmonic, log $u$ is superharmonic, so that the divergence theorem 
implies that $m_{E}(s)$ is monotone decreasing in $s$. Hence the limit 
(0.7) is well-defined, (possibly $-\infty$). Note that the static 
vacuum equations are invariant under multiplication of $u$ by positive 
constants. We use the $logu$ term in (0.7) in place of $u$ so that the 
mass is independent of this rescaling in $u$. 

  The following is the main result of this paper.

\bbgin{theorem} \label{t 0.3.}
  Let $(M, g, u)$ be a solution of the static vacuum Einstein equations.

\noindent
(i). Suppose $\partial M$ is pseudo-compact.

 Then $\bar M \setminus U$ has only finitely many ends $\{E_{i}\}, 0 
\leq  i \leq  q <  \infty .$ Supposing $q > $ 0 and (i) holds, let $E 
\in  \{E_{i}\}$ be any end of $M$ satisfying

\noindent
(ii). $u(x_{i}) \geq  u_{o},$ for some constant $u_{o} > $ 0 and some 
sequence $x_{i}\in E$ with $t(x_{i}) \rightarrow  \infty .$

 Then the end $(E, g, u)$ is either asymptotically flat, or small, in 
the sense that the area growth of geodesic spheres satisfies
\begin{equation} \label{e0.8}
\int^{\infty}\frac{1}{area S_{E}(s)}ds = \infty . 
\end{equation}
Further, if $m_{E} \neq $ 0 and $sup_{E} u <  \infty ,$ then the end 
$E$ is asymptotically flat.
\end{theorem}

 We make several remarks on this result. First, if $\partial M$ is not 
pseudo-compact, then it is easy to construct static vacuum solutions 
for which $\bar M \setminus U$ has infinitely many ends, c.f. the end 
of \S 1 or \S 2(I). Further, there are examples of ends $E$ with 
compact boundary, on which (ii) does not hold and which are neither 
asymptotically flat nor small in the sense of (0.8), c.f. Example 2.11. 
Thus, both hypotheses (i) and (ii) are necessary in Theorem 0.3. There 
are also examples with $\bar M$ compact, c.f. \S 2(I), so that $\bar M 
\setminus U$ may have no ends, ($q = 0$).

 On the other hand, both alternatives in Theorem 0.3, namely 
asymptotically flat or small ends, do occur. Asymptotically flat ends 
satisfy $area S_{E}(s) \sim 4\pi s^{2}$ as $s \rightarrow \infty$, 
while small ends have small area growth. For example, (0.8) implies, at least, 
that there is a sequence $s_{i} \rightarrow  \infty $ such that
$$area S_{E}(s_{i}) \leq  s_{i}\cdot  (log s_{i})^{1+\varepsilon}, $$
for any fixed $\varepsilon  > $ 0. All known examples of solutions 
satisfying (0.8) are topologically of the form $({\Bbb R}^{2}\setminus 
B)\times S^{1}$ outside a compact set and the geodesic spheres have at 
most linear area growth. The main example of a static vacuum solution
with a small end is the family of Kasner metrics, c.f. Example 2.11 below.

  To illustrate the sharpness of the last statement in Theorem 0.3, we 
construct in Remark 3.8 a (dipole-type) static vacuum solution $(M, g, 
u)$ with a single end $E$ on which $m_{E} =$ 0, $sup_{E} <  \infty ,$ 
and which satisfies (0.8). Similarly, Example 2.11 provides static 
vacuum solutions with $m_{E} > 0$, $sup_{E}u = \infty$ and satisfying 
(0.8). Thus, the last result is also sharp.

  Assuming $\partial M$ is pseudo-compact, it is easy to see that if 
$m_{E} > 0$ then (ii) holds, so that (ii) may be replaced by the 
assumption $m_{E} > 0$. Thus, for the physically very reasonable class 
of solutions such that $\partial M$ is pseudo-compact, $m_{E} > 0$ for 
all ends $E$, and $u$ is bounded, all ends $E$ of $M$ are 
asymptotically flat. We also point out that if $M$ is complete away 
from $\Sigma$, then (ii) holds at least on some end $E$.

  The proof of Theorem 0.3 gives some further information on the asymptotic
structure of the small ends. For instance, the curvature in such ends
decays at least quadratically, c.f. (1.3), and in the geodesic annuli
$A_{E}(\frac{1}{2}s_{i}, 2s_{i})$ in $E$, the metric approaches in a natural 
sense that of a Weyl solution, i.e. a static axisymmetric solution, as 
$s_{i} \rightarrow \infty$. Thus, asymptotically, small ends have at 
least one non-trivial Killing field. However, it is not known for example 
if the metric is asymptotic to a unique Weyl solution, or even if small
ends are necessarily of finite topological type.

  It would also be of interest to prove that $(M, g, u)$ has a unique 
end if $\partial M$ is pseudo-compact. However, we have not been able 
to do this without further assumptions, c.f. Remark 3.10.

\medskip

 The definition of asymptotically flat rules out the possibility that 
$M$ is smooth up to $\Sigma $ and complete away from $\Sigma ,$ with 
$\Sigma $ non-compact, since for example $u = 0$ on $\Sigma$ while 
$u \rightarrow 1$ on any asymptotically flat end. Consider however 
the following (B1) solution, 
\begin{equation} \label{e0.9}
g_{B1} = (1 -  \tfrac{2m}{r})^{-1}dr^{2} + (1 -  
\tfrac{2m}{r})d\phi^{2} + r^{2}d\theta^{2}, 
\end{equation}
where $\phi/4m \in [0,2\pi ], \theta\in [0,\pi ], r \geq  2m > $ 0, 
with $u = r\cdot  sin(\theta ).$ Note that the potential $u$ is 
unbounded.

 The B1 metric is just a 3-dimensional slice of the 4-dimensional 
Schwarzschild metric $(N, g_{S}+ u^{2}d\theta^{2}), (t$ changed to 
$\theta $ in (0.2)), obtained by dividing $N$ by an $S^{1}\subset 
Isom(S^{2})$ orthogonal to $d\theta ;$ this is the slice 'orthogonal' 
to the usual slice giving the Schwarzschild metric (0.4). This metric 
has $\Sigma $ given by two disjoint, isometric copies of ${\Bbb 
R}^{2},$ each of positive Gauss curvature and asymptotic to a flat 
cylinder. It is smooth up to $\Sigma$ and complete away from $\Sigma$. 
The metric is globally asymptotic to the flat metric on ${\Bbb 
R}^{2}\times S^{1},$ again with curvature decay of order $O(1/r^{3})$ 
and with $u$ of linear growth in distance to $\Sigma .$ Such solutions 
will be called {\it  asymptotically cylindrical}. In fact a large class 
of Weyl solutions have 'dual` solutions in this sense which are 
asymptotically cylindrical, c.f. Remark 2.9.

\medskip

 This paper is organized as follows. Following discussion of some 
general topics on static space-times in \S 1, we analyse in some detail 
the class of Weyl vacuum solutions in \S 2. Several new results on the 
structure of these solutions are given; for instance Proposition 2.2 
gives a new characterization of Weyl metrics. In addition, some efforts 
have been made to give a reasonably clear and organized account of the 
breadth of possibilities and behavior of Weyl metrics, since their 
treatment in the literature is rather sketchy and since they serve as a 
large class of models on which to test Theorem 0.3. Theorem 0.3 is 
proved in \S 3, and the paper concludes with several remarks on 
generalizations, and some open questions.

  I would like to thank the referee for suggesting a number of 
improvements in the exposition of the paper.

\section{Background Discussion.}
\setcounter{equation}{0}

 Let $(M, g, u)$ be an open, connected oriented Riemannian 3-manifold 
and $N = M\times_{u} {\Bbb R}$ or $N = M\times_{u} S^{1},$ as in (0.2). 
Thus $N$ represents a static space-time and $M\subset N$ is totally 
geodesic. The Einstein field equations on $N$ are
\begin{equation} \label{e1.1}
r_{N} -  \frac{s_{N}}{2}\cdot  g_{N} = T, 
\end{equation}
where $T$ is the $({\Bbb R}$ or $S^{1}$-invariant) stress-energy 
tensor. (We are ignoring physical constants here). These equations may 
be expressed on the space-like hypersurface $M$ as the system
\begin{equation} \label{e1.2}
r -  u^{-1}D^{2}u +(u^{-1}{\Delta u}-\tfrac{1}{2}s)\cdot  g = T_{H}, 
\end{equation}

$$\tfrac{1}{2}s = T_{V}, $$where $T_{H}$ is the horizontal or space-like
part of $T$ and $T_{V}$ is the vertical or time-like part of $T$. These 
are the equations on $M$ for a Lorentzian space-time $N$; in case $N$ 
is Riemannian, the first equation is the same while the second is 
-$\frac{1}{2}s = T_{V}$. When $T =$
0, one obtains the vacuum equations (0.1), which are the same for 
Lorentzian or Riemannian signature. A common example, with $T \neq 0$ 
is a {\it static perfect fluid}, c.f. [Wd, Ch.4], with $T$ given by
$$T = (\mu +\rho )dt^{2} + \rho g, $$where $\mu , \rho $ are time 
independent scalar fields representing the energy density and pressure 
respectively. The equations (1.2) imply that the full Riemann curvature 
$R_{N}$ of $N$ is determined by $r$, $u$ and $T$.

 The horizon $\Sigma  = \{u=0\}$ corresponds formally to the fixed 
point set of the $S^{1}$ action on $N$ and requires special 
consideration. For example, even if $M$ is smooth up to $\Sigma $ the 
Riemannian 4-manifold $(N, g_{N})$ might not be smooth across $\Sigma 
,$ even though the curvature $R_{N}$ is smooth. Namely, assuming the 
$S^{1}$ parameter $t\in [0,2\pi )$, if $|\nabla u|_{\Sigma} \neq $ 1, 
then $N$ has cone singularities (with constant angle by (0.3)) along 
and normal to the totally geodesic submanifold $\Sigma\subset N.$ (This 
issue does not arise for Lorentzian metrics). By multiplying the 
potential function $u$ by a suitable constant, one can make the metric 
$g_{N}$ smooth across any given component of $\Sigma ;$ one cannot 
expect however in general that this can be done simultaneously for all 
components of $\Sigma ,$ if there are more than one. This issue will 
reappear in \S 2.

\medskip

 The following result is proved in [A1, Cor.A.3]. It implies, (by 
letting $t\rightarrow\infty ),$ that if $(M, g, u)$ is a complete 
solution to the static vacuum equations with $u > $ 0 everywhere, then 
$M$ is flat, and $u$ is constant.

\bbgin{theorem} \label{t 1.1.}
  Let (M, g, u) be a solution to static vacuum equations (0.1). Let 
t(x) $=$ dist(x, $\partial M)$ as in (0.6). Then there is a constant $K 
<  \infty ,$ independent of (M, g, u), such that
\begin{equation} \label{e1.3}
|r|(x) \leq  \frac{K}{t(x)^{2}}, |\nabla logu|(x) \leq  \frac{K}{t(x)}. 
\end{equation}
\end{theorem}

\bbgin{remark} \label{Remark 1.2.(i)}
{\rm The same result has recently been proved for stationary vacuum 
solutions, i.e. space-times admitting a complete time-like Killing 
field, c.f. [A2]. We will discuss elsewhere to what extent Theorem 0.3 
generalizes to stationary vacuum solutions.}

\end{remark}
\medskip
 Recall from \S 0 that the potential function $u$ of a static vacuum 
solution may be freely renormalized by arbitrary positive constants; 
hence the appearance of log $u$ in (1.3), as in (0.7).

 Theorem 1.1 implies that the curvature of $(M, g)$ is controlled away 
from $\partial M,$ and hence the local geometry of solutions is 
controlled away from $\partial M$ by lower bounds on the local volume 
or injectivity radius. More precisely, we have the following results 
which are essentially a standard application of the Cheeger-Gromov 
theory of convergence/collapse of Riemannian manifolds, c.f. [CG], [A3, 
\S2], or also [P, Ch.12] for an introduction to these results.
Further details of the proofs of these results are given in [A2], 
(for the more general class of stationary space-times), and also in 
[A1,App.].

\bbgin{lemma} \label{l 1.3}
 {\bf (Non-Collapse).}
 Let $(M_{i}, g_{i}, u_{i})$ be a sequence of solutions to the static 
vacuum equations (0.1). Suppose
$$
|r_{i}| \leq  \Lambda , diam M_{i} \leq  D, vol M_{i} \geq  \nu_{o}, 
$$and
$$dist(x_{i}, \partial M_{i}) \geq  \delta , $$
for some $x_{i}\in M_{i}$ and positive constants $\nu_{o}, \Lambda ,$ 
D, $\delta .$ Assume also that $u_{i}$ is normalized so that 
$u_{i}(x_{i})$ = 1. 

Then, for any $\epsilon  > $ 0 sufficiently small, there are domains 
$U_{i}\subset M_{i},$ with $\epsilon /2 \leq  dist(\partial U_{i}, 
\partial M_{i}) \leq  \epsilon ,$ and $x_{i}\in U_{i}$ such that a 
subsequence of the Riemannian manifolds $(U_{i}, g_{i})$ converges, in 
the $C^{\infty}$ topology, modulo diffeomorphisms, to a limit manifold 
(U, g), with limit function $u$ and base point $x =$ lim $x_{i}.$ The 
triple (U, g, u) is a smooth solution of the static equations with u(x) 
$=$ 1.
\end{lemma}

\bbgin{lemma} \label{l 1.4}
{\bf (Collapse).}
 Let $(M_{i}, g_{i}, u_{i})$ be a sequence of solutions to the static 
vacuum equations (0.1). Suppose
$$
|r_{i}| \leq  \Lambda , diam M_{i} \leq  D, vol M_{i} \rightarrow  0 
$$and
$$dist(x_{i}, \partial M_{i}) \geq  \delta , $$
for some $x_{i}\in M_{i}$ and constants $\Lambda ,$ D, $\delta .$ 
Assume also that $u_{i}$ is normalized so that $u_{i}(x_{i})$ = 1. 

Then, for any $\epsilon  > $ 0 sufficiently small, there are domains 
$U_{i}\subset M_{i},$ with $\epsilon /2 \leq  dist(\partial U_{i}, 
\partial M_{i}) \leq  \epsilon $ with $x_{i}\in U_{i},$ such that 
$U_{i}$ is either a Seifert fibered space or a torus bundle over an 
interval. In both cases, the $g_{i}$-diameter of any fiber F, 
(necessarily a circle $S^{1}$ or torus $T^{2}),$ goes to 0 as $i 
\rightarrow  \infty ,$ and $\pi_{1}(F)$ injects in $\pi_{1}(U_{i}).$

 Consequently, there are infinite ${\Bbb Z}$ or ${\Bbb Z} \oplus {\Bbb 
Z}$ covers $\Roof{U}{\widetilde}_{i}$ of $U_{i}$, such that 
$\{\widetilde U_i, g_i, x_i\}$ does not collapse and hence has a 
subsequence converging smoothly to a limit $(\Roof{U}{\widetilde}, g, 
x)$ of the static vacuum equations with $x =$ lim $x_{i}' , x_{i}' $ a 
lift of $x_{i}$ to $\Roof{U}{\widetilde}_{i}.$ The limit 
$(\Roof{U}{\widetilde}, g, x)$ admits a free isometric ${\Bbb R}$ or 
${\Bbb R} \oplus {\Bbb R}$ action, (c.f. \S 2), which also leaves the 
potential function $u$ invariant, and $u(x) =$ 1.
\end{lemma}

\medskip
 In studying static solutions, it is often very useful to consider the 
conformally equivalent metric $\Roof{g}{\tilde} = u^{2}\cdot  g$ on 
$M$. An easy calculation using the behavior of Ricci curvature under 
conformal deformations, c.f. [Bes, p.59], shows that the Ricci 
curvature $\Roof{r}{\tilde}$ of $\Roof{g}{\tilde},$ in the vacuum case 
(0.1), is given by
\begin{equation} \label{e1.4}
\Roof{r}{\tilde} = 2(dlogu)^{2} \geq  0. 
\end{equation}
Further, if $\Roof{\Delta}{\tilde}$ denotes the Laplacian of 
$\Roof{g}{\tilde},$ then
\begin{equation} \label{e1.5}
\Roof{\Delta}{\tilde}logu = 0. 
\end{equation}
The equations (1.4)-(1.5) are equivalent to the static vacuum equations 
(0.1). Since these equations are invariant under the substitution $u 
\rightarrow  - u,$ it follows that if $(M, g, u)$ is a static vacuum 
solution, then so is $(M, g' , u^{-1}),$ with $g' $ given by
$$g'  = u^{4}\cdot  g. $$

 Similarly, observe that if $(N, g_{N})$ is the associated Ricci-flat 
4-manifold (0.2), then
\begin{equation} \label{e1.6}
\Delta_{N}logu = 0. 
\end{equation}
Here and below, log always denotes the natural logarithm.

 We discuss briefly some of the simplest static vacuum solutions:

\medskip

{\bf Levi-Civita Solutions.}

 There are 7 classes of so-called degenerate static vacuum solutions, 
where the eigenvalues $\lambda_{i}$ of the Ricci curvature $r$ satisfy 
$\lambda_{1}=\lambda_{2}=- 2\lambda_{3},$ called A1-A3, B1-B3, C, c.f. 
[EK, \S 2-3.6]. The $B$ metrics are dual to the $A$ metrics, as 
mentioned in \S 0, c.f. \S 2 for details. The A1 metric is the 
Schwarzschild metric. It is of interest to examine the A2 metric, given 
in standard cylindrical coordinates on ${\Bbb R}^{3}$ by
\begin{equation} \label{e1.7}
g_{A2} = z^{2}(dr^{2}+(sinh^{2}r)d\phi^{2}) + (\frac{2m}{z}- 
1)^{-1}dz^{2}, 
\end{equation}
with $u = (\frac{2m}{z}- 1)^{1/2}$ and $z\in [0,2m], m > 0$. The 
horizon $\Sigma  = \{u=0\}$ is given by the set $\{z = 2m\}$ and hence 
is the complete hyperbolic metric $H^{2},$ with curvature $-(2m)^{-2}.$ 
It is easily verified that the A2 metric is smooth up to $\Sigma .$ 
However, $\Sigma  \neq  \partial M;$ the set $\{z =$ 0\} is at finite 
distance to $\Sigma ,$ and so $\partial M$ has another (singular) 
component obtained by crushing (compact subsets of) the hyperbolic 
metric to a point.

 Let $\Gamma $ be any discrete group of hyperbolic isometries. Then 
$\Gamma $ extends in an obvious way to a group of isometries of 
$g_{A2}.$ The uniformization theorem for surfaces implies that any 
orientable surface except $S^{2}$ and $T^{2},$ including surfaces of 
infinite topological type and infinitely many ends, admits a complete 
hyperbolic metric, i.e. is the quotient $H^{2}/\Gamma ,$ for some 
$\Gamma .$ Hence, any such surface and hyperbolic metric can be 
realized as the horizon $\Sigma $ of a static vacuum solution.

  Topologically, for $\Sigma = H^{2}/ \Gamma$, we have $M$ = 
$\Sigma\times I$. Hence for example if $\Sigma$ has infinitely many 
ends, then $M$ also has infinitely many ends; in particular, this shows 
that the hypothesis that $\partial M$ is pseudo-compact in Theorem 0.3 
is necessary.

 The A3 metric is
\begin{equation} \label{e1.8}
g_{A3} = z^{2}(dr^{2}+r^{2}d\phi^{2}) + zdz^{2}, 
\end{equation}
with $u = z^{-1/2} > 0, r > 0.$ Hence $\Sigma $ is empty in this case - 
it occurs at infinity in the metric. This metric may be realized as a 
pointed limit of the A2 metric as $m \rightarrow  \infty $ and also as 
a limit, in a certain sense, of the A1 metric, c.f. Example 2.11.

\medskip
\bbgin{remark} \label{Remark 1.5}
 {\rm The discussion above raises the natural question if any 
orientable connected Riemannian surface $(\Sigma , g)$ can be realized 
as the horizon of a static vacuum solution, smooth up to $\Sigma ,$ 
which is defined at least in a neighborhood of $\Sigma .$ In general, 
this appears to be unknown.

 Observe that any complete constant curvature metric on an orientable 
surface can be realized in this way. The Schwarzschild metric gives the 
constant curvature metric on $S^{2},$ the quotients of the A2 metric 
give all hyperbolic surfaces, and quotients of the flat metric, with 
$u$ a linear function, give all flat metrics on a surface, $(T^{2}, 
S^{1} \times {\Bbb R}, {\rm or} \ {\Bbb R}^{2}).$

 Geroch-Hartle show in [GH] that any rotation-invariant metric on 
$S^{2}$ or $T^{2}$ can be realized at the horizon. Except for the 
Schwarzschild metric, such solutions are not complete away from $\Sigma 
.$ 

 Observe that the full 1-jet of $(M, g)$ at $\Sigma $ (assumed 
connected) is determined solely by the surface metric $(\Sigma , g)$, 
since $\Sigma $ is totally geodesic and, renormalizing $u$ if 
necessary, $|\nabla u| \equiv $ 1 on $\Sigma $ by (0.3). Observe also 
that one cannot have $|\nabla u| \equiv $ 0 on $\Sigma ,$ since $u$ is 
harmonic and the divergence theorem applied to a small neighborhood $U$ 
of $\Sigma $ would imply that $u \equiv $ 0 on $U$, which is ruled out.

 On the other hand, the metric $(M, g)$ is not uniquely determined by 
its boundary values $(\Sigma , g)$. Namely, the flat metric on $T^{2}$ 
is realized by the flat vacuum solution $M = T^{2}\times {\Bbb R}^{+}, 
u = t = dist(\Sigma , \cdot  )$ and also (locally) by a non-flat 
metric, c.f. [T],[P]. Similar remarks hold for local perturbations of 
the Schwarzschild metric, c.f. [GH].}
\end{remark}

\section{Weyl Solutions.}
\setcounter{equation}{0}

 A large and very interesting class of explicit solutions of the static 
vacuum equations are given by the {\it  Weyl solutions}  [W], c.f. also 
[EK,\S 2.3-9] or [Kr, Ch.16-18]. In fact, it appears that essentially 
all known explicit solutions of the static vacuum equations are of this 
form. Since the literature on these solutions is not very organized or 
rigorous, especially regarding their global structure, we discuss these 
solutions in some detail. These metrics will also illustrate the 
necessity of the hypotheses in Theorem 0.3.

\medskip
\bbgin{definition} \label{Definition 2.1.}
  A Weyl solution is a solution $(M, g, u)$ of the static vacuum 
equations (0.1) which admits an isometric ${\Bbb R}$-action, i.e. a 
non-zero homomorphism ${\Bbb R}  \rightarrow Isom(M)$, leaving $u$ 
invariant.
\end{definition}
\medskip
 Apriori, the topology of a Weyl solution could be quite non-trivial; 
for example $M$ could be any Seifert fibered space. The first result 
shows that only the simplest topology (and geometry) is possible. For 
the moment, we exclude any possible fixed point set of the ${\Bbb 
R}$-action from the discussion.

\bbgin{proposition} \label{p 2.2.}
  Let $(M, g, u)$ be a Weyl solution with ${\Bbb R}$-action without 
fixed points, which does not admit a (local) free isometric ${\Bbb R} 
\times {\Bbb R}$ action. Then the universal cover $(\widetilde M, g)$ 
of $(M, g)$ is a warped product of the form
\begin{equation} \label{e2.1}
\widetilde M = V\times_{f}{\Bbb R}, g = g_{V} + f^{2}d\phi^{2}, 
\end{equation}
with (V, $g_{V})$ a Riemannian surface and $f$ a positive function on 
V. The ${\Bbb R}$-action on $\widetilde M$ is by translation on the 
second factor.
\end{proposition}
{\bf Proof:}
  This result, whose proof is purely local, is a strengthening in this 
situation of a well-known result in general relativity, Papapetrou's 
theorem, c.f. [Wd,Thm.7.1.1], which requires certain global 
assumptions, (e.g. smoothness up to $\Sigma ).$ 

  Let $K$ denote the (complete) Killing field generated by the ${\Bbb 
R}$-action on $M$, and $f = |K|$. We may assume that $u$ is not a 
constant function on $M$, since if $u$ is constant, the metric is flat, 
and so admits a local ${\Bbb R} \times {\Bbb R}$ action. Since $(M, g, 
u)$ is real-analytic, $u$ is not constant on any open set in $M$. We 
thus choose a neighborhood $U$ of any point $p$ where $\nabla u(p) \neq 
0$ on which $|\nabla u| > 0$. Define $e_{1}$ by $e_{1} = \nabla u 
/|\nabla u|$, and extend it to a local orthonormal frame $e_{1}, e_{2}, 
e_{3}$ for which $e_{3} = K/|K| = K/f$. Note that this is possible 
since $u$ is required to be invariant under the flow of $K$, so that
\begin{equation} \label{e2.2}
<\nabla u, K> = 0. 
\end{equation}

 In $U$, the metric $g$ may be written as
\begin{equation} \label{e2.3}
g = \pi^{*}g_{V} + f^{2}(d\phi  + \theta )^{2}, 
\end{equation}
where $\pi : U \rightarrow  V$ is a Riemannian submersion onto a local 
surface $(V, g_{V})$, $\theta $ is a connection 1-form, $K = \partial 
/\partial\phi $ and $f$ is a function on the orbit space $V$. If 
$\theta  =$ 0, then the result follows. Thus, we assume $|\theta| > $ 0 
in $U$ and show this implies that $g$ has a free isometric local ${\Bbb 
R} \times {\Bbb R} $ action.

 Consider the 1-parameter family of metrics
\begin{equation} \label{e2.4}
g_{s} = \pi^{*}g_{V} + s^{2}f^{2}(d\phi  + s^{-2}\theta )^{2}, 
\end{equation}
for $s > $ 0, with $g_{1} = g$. Geometrically, this corresponds to 
rescaling the length of the fibers of $\pi $ and changing the 
horizontal distribution of $\pi ,$ (when $\theta  \neq $ 0). Now it is 
a standard fact that the 1-parameter family of 4-metrics
\begin{equation} \label{e2.5}
g_{s}^{4} = g_{s} \pm  u^{2}dt^{2} 
\end{equation}
remains Ricci-flat for all $s$. This can be seen from standard formulas 
for Riemannian line bundles, c.f. [Bes, 9.36, 9G], [Kr, 16.1-3] or [A2, 
\S 1.2]. Thus, the metrics $g_{s}$ all satisfy the static vacuum 
equations
$$ur_{s} = D_{s}^{2}u $$
from (0.1), with the same potential $u$. Equivalently, the conformal 
metrics $\tilde g_{s} = u^{2}g_{s}$ satisfy
\begin{equation} \label{e2.6}
\tilde r_{s} = 2(dlog u)^{2}, 
\end{equation}
for all $s$, c.f. (1.4) The right side of (2.6) is of course 
independent of $s$.

 We claim that the metrics $g_{s}$ on $U$ are all locally isometric. 
While this could be proved by a lengthy direct computation, we argue 
more conceptually as follows. Let $e_{i}^{s}$ be a local orthonormal 
frame for $g_{s},$ determined as above for $g$. We then have $e_{1}^{s} 
= e_{1}, e_{3}^{s} = s^{-1}e_{3}$ while $e_{2}^{s}$ varies in the plane 
$< e_{2}, e_{3}> .$ The same relations hold w.r.t. $\tilde g_{s}.$ 
Recall also that the full curvature tensor is determined by the Ricci 
curvature in dimension 3. It then follows from these remarks and (2.6) 
that for each $q\in U$ and $s > $ 0, there is a sectional curvature 
preserving isomorphism $F_{s}: T_{q}M \rightarrow  T_{q}M,$ i.e.
$$\tilde K_{s}(F_{s}(P)) = \tilde K_{1}(P), $$
where $P$ is any 2-plane and $\tilde K_{s}$ is the sectional curvature 
w.r.t. $\tilde g_{s}.$ Clearly $F_{s}$ varies smoothly with $q$ and 
$s$. Using the expression (2.6), a result of Kulkarni [Ku] then implies 
that the metrics $\tilde g_{s}$ are locally isometric and hence so are 
the metrics $g_{s}.$

 Let $\Omega  = d\theta $ be the curvature form of the line bundle $\pi 
.$  Then w.r.t. the metric $g$, $|\Omega| = |\Omega (e_{1},e_{2})| = 
|<\nabla_{e_{1}}e_{2}, e_{3}>|.$ The same equalities hold w.r.t. 
$\Omega_{s} = d\theta_{s} = s^{-2}d\theta $ and the $g_{s}$ metric. A 
short computation then gives
\begin{equation} \label{e2.7}
|\Omega_{s}|_{g_{s}} \rightarrow  0, \ \ {\rm as} \ \  s \rightarrow  
\infty . 
\end{equation}

 Hence consider the behavior of the metrics $g_{s}$ as $s \rightarrow  
\infty .$ We are then expanding or blowing up the metric in the fiber 
direction, at a given base point. Since the metrics $g_{s}$ are 
isometric, there are (local) diffeomorphisms $\psi_{s}$ such that 
$\psi_{s}^{*}g_{s}$ converges to a limit metric $g_{\infty}.$ At a 
given base point, the diffeomorphisms $\psi_{s}$ expand or blow up 
smaller and smaller intervals of the parameter $\phi $ to unit size, 
giving rise to a limit parameter $\phi_{\infty}.$ When $\theta  =$ 0, 
this is the only change; the limit metric $g_{\infty}$ is the same as 
$g$ with the parameter $\phi $ replaced by $\phi_{\infty}.$ (This is 
completely analogous to passing from the flat metric on ${\Bbb R}^{2} 
\setminus \{0\}$ to the flat metric on its universal cover, i.e. 
unwrapping the circles to lines). 

 However, when $\theta  \neq $ 0, the $e_{2}$ direction is also being 
expanded or blown-up in a similar way. The function $u$ is left 
invariant under the family $\{\psi_{s}\}.$ 

 It follows from (2.7) that the limit metric $g_{\infty}$ is a static 
vacuum solution of the form
\begin{equation} \label{e2.8}
g_{\infty} = \pi^{*}g_{V_{\infty}} + 
f_{\infty}^{2}(d\phi_{\infty})^{2}, 
\end{equation}
i.e. the 1-form $\theta_{\infty} =$ 0 in the limit. Further, since the 
$e_{2}$ direction has been blown up, the function $f_{\infty}$ varies 
only in the $e_{1}$ direction, i.e. $f_{\infty} = f_{\infty}(u).$

 Metrics of the form (2.8) are analysed in detail below. Referring to 
(2.10), let $r = f_{\infty}\cdot  u = h(u)$. In a possibly smaller open 
subset of $U$, we may invert $h$ and write $u = u(r)$, where $r$ is a 
local coordinate on $V_{\infty}.$ It is easy to see, (c.f. 
(2.12)-(2.13) below for example), that $g_{\infty}$ admits a 
non-vanishing Killing field $\partial /\partial z,$ tangent to 
$V_{\infty}$ but orthogonal to $\partial /\partial r,$ and hence 
$g_{\infty}$ admits a free isometric local ${\Bbb R} \times {\Bbb R} $ 
action. The metrics $g_{s}$ are all locally isometric and so the metric 
$g = g_{1}$ also has a free isometric local ${\Bbb R} \times {\Bbb R} $ 
action.

{\endproof} 

 Since the proof above is completely local, Proposition 2.2 holds 
locally, (in suitably modified form), even if $(M, g)$ admits only a 
local or partial ${\Bbb R}$ -action.

 Static vacuum solutions admitting a free isometric local ${\Bbb R} 
\times {\Bbb R} $ action are completely classified; they are either 
flat or belong to the family of Kasner metrics, c.f. Example 2.11 below 
or [EK, Thm.2-3.12]. Such solutions {\it do} have Killing fields $K$ 
which are not hypersurface orthogonal, i.e. $d\theta  \neq $ 0 in 
(2.3). For example, if $\partial /\partial\psi $ and $\partial 
/\partial z$ are standard generators of the (local) ${\Bbb R} \times 
{\Bbb R} $ action, then linear combinations such as $K = \partial 
/\partial\psi  + \partial /\partial z$ are non-hypersurface orthogonal 
Killing fields.

 Nevertheless, all such solutions do admit, of course, hypersurface 
orthogonal Killing fields and so may be written in the form (2.1). For 
the remainder of the paper, we thus assume that a Weyl solution has the 
form (2.1) locally. In addition, we will always work with the ${\Bbb 
Z}$-quotient of the metric (2.1) and so consider Weyl solutions as 
warped products of the form $V \times_{f} S^{1}$; it will not be 
assumed in general that $V$ is simply connected.

\medskip

{\bf Duality.}
 Observe that Weyl solutions naturally come in 'dual' pairs. Namely the 
Ricci-flat 4-manifold $(N,g_{N})$ has the form
\begin{equation} \label{e2.9}
g_{N} = g_{V} + f^{2}d\phi^{2} + u^{2}dt^{2}, 
\end{equation}
and so both $M_{u}=V\times_{f}S^{1}$ and $M_{f}=V\times_{u}S^{1}$ are 
static vacuum solutions on the 3-manifolds, with potentials $u$, resp. 
$f$.
 Consider the product of the lengths of circles, or equivalently, the 
area of the torus fiber in $N$,
\begin{equation} \label{e2.10}
r = f\cdot  u .
\end{equation}
This is a globally defined positive {\it harmonic}  function on $(V, 
g_{V}).$ To see this, on $M = M_{u},$ by (2.5), we have
$$0 = \Delta u = \Delta_{V}u  + <\nabla logf, \nabla u> , $$and the 
same formula, with $u$ and $f$ reversed, holds on $M_{f}.$ Hence
$$\Delta_{V}fu = f\Delta_{V}u + u\Delta_{V}f + 2<\nabla u, \nabla f>  = 
0. $$

{\bf Charts.}
 We now describe a collection of preferred charts in which to express 
the Weyl solution $(M, g)$; this description is due to Weyl [Wl]. The 
surface $V$ may be partitioned into a collection of maximal domains 
$V_{i}$ on which the harmonic conjugate $z$ of $r$ is single-valued, so 
that $F = r+iz$ is a well-defined holomorphic function from $V_{i}$ 
into the right half-plane ${\Bbb C}^{+} =$ \{$(r,z)$: $r > $ 0, 
$z\in{\Bbb R}\}.$ One might also pass to a suitable cover, for instance 
the universal cover, of $V$ to obtain a globally defined conjugate 
harmonic function, but it is preferable not to do so.

 Now each $V_{i}$ may be further partitioned into a collection of 
domains $V_{ij}$ on which $F$ is a conformal embedding into ${\Bbb 
C}^{+},$ so that $g|_{V_{ij}} = F^{*}(dr^{2}+dz^{2}).$ We will thus 
simply view $V_{ij}$ as a domain in ${\Bbb C}^{+},$ with $g_{V}$ a 
metric pointwise conformal to the flat metric $dr^{2}+dz^{2}.$

 It follows that the corresponding domain $M_{ij} = 
V_{ij}\times_{f}S^{1}$ is embedded as a domain $\Omega  = \Omega_{ij}$ 
in ${\Bbb R}^{3}$ endowed with cylindrical coordinates $(r,z,\phi)$, 
$\phi\in [0,2\pi )$ with the background (unphysical) complete flat 
metric $dr^{2}+dz^{2}+r^{2}d\phi^{2}.$ We note that all the data above 
are canonically determined by the two Killing fields on $N$ and thus 
the coordinates $(r, z, \phi )$ are called canonical cylindrical or 
Weyl coordinates for $(M, g)$. Of course $\Omega $ is axially 
symmetric, i.e. symmetric w.r.t. rotation about the $z$-axis.

 To express the metric $g|_{\Omega}$ in these coordinates, the field 
equations (0.1) imply that the function
\begin{equation} \label{e2.11}
\nu  = log u 
\end{equation}
is an axially symmetric (independent of $\phi )$ {\it harmonic}  
function on $\Omega  \subset  {\Bbb R}^{3};$ this again follows in a 
straightforward way from computation of the Laplacian of $u$ and $f$ on 
$M_{f}$ and $M_{u}$ as above. A computation of the conformal factor for 
the metric $g_{V},$ c.f. [Wd,Ch.7.1], then leads to the expression of 
$g$ in these coordinates:
\begin{equation} \label{e2.12}
g = u^{-2}(e^{2\lambda}(dr^{2}+dz^{2}) + r^{2}d\phi^{2}), 
\end{equation}
where $\lambda $ is determined by $\nu $ as a solution to the 
integrability equations
\begin{equation} \label{e2.13}
\lambda_{r} = r(\nu_{r}^{2}-\nu_{z}^{2}), \lambda_{z} = 
2r\nu_{r}\nu_{z}. 
\end{equation}
The equations (2.13) mean that the 1-form $\omega  = 
r(\nu_{r}^{2}-\nu_{z}^{2})dr + 2r\nu_{r}\nu_{z}dz$ is closed on $\Omega 
.$

 Conversely, given any axially symmetric harmonic function $\nu $ on a 
connected open set $\Omega $ in ${\Bbb R}^{3},$ if the closed 1-form 
$\omega $ is exact, (for example if $\pi_{1}(\Omega\cap{\Bbb 
C}^{+})=0),$ the equations (2.13) determine $\lambda $ up to a constant 
and the metric (2.12) gives a solution to the static vacuum equations 
with $S^{1}$ symmetry. (The addition of constants to $\nu $ or $\lambda 
$ changes the metric at most by diffeomorphism or homothety). 

 It is remarkable that solutions to the non-linear vacuum equations 
(0.1) can be generated in this way by solutions to the linear Laplace 
equation on ${\Bbb R}^{3}.$

\medskip
\bbgin{remark} \label{Remark 2.3.}
 {\bf (i).}  {\rm The Levi-Civita solutions in \S 1 are all Weyl 
solutions. However, the expressions for the A1-A3 metrics in 
(0.4),(1.7),(1.8) and the B1 metric in (0.9) are not in Weyl canonical 
coordinates. Note that the quotients of the A2 metric discussed above 
are no longer Weyl solutions, although they could be considered as 
local Weyl solutions; the ${\Bbb R}$-action is only locally defined on 
the quotients.

{\bf (ii).}
 There seem to be no known Weyl solutions which can not be expressed 
globally in the form (2.12).}
\end{remark}
\medskip
{\bf Fixed Point Set.}
 The behavior of solutions at the part of $\partial M$ where either one 
of the two $S^{1}$ or ${\Bbb R}$ actions on $N$ has fixed points 
requires special considerations. This is of course the locus where $u 
=$ 0 or $f =$ 0, and hence includes the part $\Roof{\Omega}{\bar}\cap 
A$ of the $z$-axis $A = \{r=0\}$ in any canonical coordinate chart. It 
is not necessarily the case however that this locus is contained in 
$A$, c.f. the end of Example 2.10.

 Note that given any Weyl solution (2.12), any covering of ${\Bbb 
R}^{3}\setminus A$ induces another solution of the form (2.12), but 
with $\phi $ parametrizing a circle of length $2\pi k.$ For the 
universal cover $(k = \infty ),$ the $\phi$-circle is replaced by a 
line. In fact, (2.12) is well-defined when $\phi $ runs over any 
parameter interval [0, $2\pi\alpha ).$ Observe however that any 
asymptotically flat Weyl solution must have $\alpha  =$ 1, since the 
metric must be smooth near infinity. Thus, we will assume $\alpha  =$ 1 
in the following, unless stated otherwise.

 Now suppose there is an open interval $J$ in $A$ such that the 
functions $u$ and $\lambda ,$ and hence the form (2.12) extend 
continuously to $J$. The form $g$ then represents a continuous metric 
in a neighborhood of $J$ if and only if the {\it  elementary flatness 
condition}
\begin{equation} \label{e2.14}
\lambda  = 0, 
\end{equation}
is satisfied on $J$. On intervals where (2.14) does not hold, the 
metric $g$ has cone singularities, so that it is not locally Euclidean. 
From (2.13), it is clear that if $\lambda $ has a $C^{1}$ extension to 
$J$, then $\lambda $ is constant on $J$. However, such constants may 
vary over differing components of $\Roof{\Omega}{\bar}\cap A.$ This 
will be analysed further in Remark 2.8.

\medskip

 For the remainder of \S 2, we assume that 
$$M = \Omega , $$so that the Weyl solution is given globally in the 
form (2.12). Let $I$ be the set where $\nu =-\infty ,$ i.e. the 
$G_{\delta}$ set in $\Omega\subset{\Bbb R}^{3}$ given by 
\begin{equation} \label{e2.15}
I = \bigcap _{n}\nu^{-1}(-\infty , n), 
\end{equation}
where $n$ runs over negative integers. It is usually assumed in physics 
that $I$ is non-empty, although this need not be the case; this will 
also be discussed further below. The set $I$ corresponds to the horizon 
$\Sigma $ of the Weyl solution $(M, g)$, since $u =$ 0 on $I$. This 
correspondence is formal however, since the geometry and topology of 
$I\subset{\Bbb R}^{3}$ is very different than that of $\Sigma\subset 
(\Roof{M}{\bar}, g)$, c.f. most of the examples below. For the same 
reasons, although $M = \Omega $ topologically, the metric boundary 
$\partial M$ of $(M, g)$ is (most always) very different than the 
Euclidean boundary of $\Omega\subset{\Bbb R}^{3}.$

 In the following, we discuss some of the most significant possible 
behaviors for the potential function $u$, and the associated Weyl 
solution, in order to illustrate the breadth of these solutions. The 
discussion is by no means complete or exhaustive.

\medskip

{\bf (I). $\Roof{\Omega}{\bar}$ compact.}

 Let $M = \Omega $ be any bounded, $C^{\infty}$ smooth axisymmetric 
domain (i.e. connected open set with smooth compact boundary) in ${\Bbb 
R}^{3}$ and let $\phi$ be any $C^{k,\alpha}$ function on 
$\partial\Omega ,  k \geq $ 0, $\alpha  \in $ (0,1), which is axially 
symmetric about the $z$-axis. For simplicity, assume that 
$\Omega\cap{\Bbb C}^{+}$ is simply connected. Let $\nu $ be the 
solution to the Dirichlet problem
$$\Delta\nu  = 0, \nu|_{\partial\Omega} = \phi. $$Then $\nu $ is also 
axi-symmetric about the $z$-axis, and hence $\nu $ generates a Weyl 
solution as in (2.12). 

 Suppose that for a given $k \geq $ 1 $\alpha  \in $ (0,1), $\phi$ as 
above is $C^{k,\alpha}$ on $\partial\Omega ,$ but is nowhere $C^{k+1}$ 
on $\partial\Omega .$ Then $\nu $ extends to a $C^{k,\alpha}$ function 
on the Euclidean closure $\Roof{\Omega}{\bar}$ and hence, from (2.13), 
the function $\lambda $ in (2.12) is also uniformly bounded. This means 
that the metric $g$ is quasi-isometric to the flat metric on $\Omega ,$ 
and hence the metric boundary of $\Omega $ w.r.t.  the Weyl metric $g$ 
is the same as its Euclidean boundary. Since $\nu $ is not $C^{k+1}$ 
anywhere on $\partial\Omega ,$ this solution $M = \Omega $ is maximal, 
i.e. admits no larger static vacuum extension; $C^{2}$ smooth solutions 
of the static vacuum equations are analytic. Further $\nu $ is bounded, 
so that $u = e^{\nu}$ is bounded away from 0, and hence the solution 
$(\Omega , g)$ has no horizon. 

 As noted in \S 0, the presence of the boundary $\partial M$ is 
physically assumed due to the presence of matter or field sources. 
Thus, at least when $k \geq $ 2, the vacuum solution $(M, g)$ can be 
extended to a larger space-like domain $(M', g')\supset (M, g)$ with 
non-zero stress-energy $T$ in $M' \setminus M.$

 On the other hand, if $k =$ 0 above, then the geometry of the metric 
boundary $(\partial M, g)$ will in general be very different than the 
smooth geometry of $\partial\Omega $ in ${\Bbb R}^{3}.$ Further, one 
can of course consider non-smooth domains $\Omega \subset {\Bbb R}^{3}$ 
in this situation. These remarks indicate that the structure of the 
metric boundary $\partial M$ seemingly can be quite arbitrary.

\medskip

 For the remainder of this section, we assume that 
$\Roof{\Omega}{\bar}$ is non-compact in ${\Bbb R}^{3}.$ The same 
remarks as above hold for non-compact domains with smooth (surface) 
boundary. Thus for example $M = \Omega$ might have infinitely many ends 
if $\partial \Omega$ is non-compact, showing that the assumption (i) in 
Theorem 0.3 is necessary. For simplicity, we only consider the 
following situation from now on.

{\bf (II).}
 Suppose
\begin{equation} \label{e2.16}
dim_{{\cal H}}\partial\Omega  \leq  1, 
\end{equation}
where the boundary is in the topology of ${\Bbb R}^{3}$ and $dim_{{\cal 
H}}$ is the Hausdorff dimension. Thus $\partial\Omega $ is a closed set 
of capacity 0, c.f. [H, Thm.5.14], and so in particular is a polar set. 
Clearly $\Roof{\Omega}{\bar} = {\Bbb R}^{3}.$

{\bf (A).(Positive Case). }
 Suppose that $\nu $ is locally bounded above, i.e.
\begin{equation} \label{e2.17}
\sup_{B_{x}(1)}\nu  <  \infty , \forall x\in\partial\Omega . 
\end{equation}

 It follows, c.f. [H, Thm.5.18] that $\nu $ extends uniquely to a 
globally defined subharmonic function on ${\Bbb R}^{3}.$ Hence, one may 
use the value distribution theory of subharmonic functions on ${\Bbb 
R}^{3}$ to analyse the geometry of Weyl solutions.

 The Riesz representation theorem c.f. [H,Thm.3.9], implies that any 
subharmonic function $\nu $ on ${\Bbb R}^{3}$ may be represented 
semi-globally, i.e. on $B(R)= B_{0}(R)\subset{\Bbb R}^{3}$ for any $R < 
\infty$, as
\begin{equation} \label{e2.18}
\nu (x) = -\int_{B(R)}\frac{1}{|x-\xi|}d\mu_{\xi} + h(x), 
\end{equation}
where $d\mu_{\xi}$ is a {\it positive}  Radon measure on $B(R)$ called 
the {\it  Riesz measure}  of $\nu $ and $h$ is a harmonic function on 
$B(R)$; both $d\mu $ and $h$ are axi-symmetric if $\nu $ is. (A Radon 
measure is a Borel measure which is finite on compact subsets). For the 
moment, we only consider the situation where there exists $K <  \infty 
,$ independent of $R$, s.t.
\begin{equation} \label{e2.19}
 \int_{B(R)}\frac{1}{|x-\xi|}d\mu_{\xi} \leq  K .
\end{equation}
In this case, one obtains a global representation of $\nu $ as
\begin{equation} \label{e2.20}
\nu (x) = -\int_{{\Bbb R}^{3}}\frac{1}{|x-\xi|}d\mu_{\xi} + h(x), 
\end{equation}
where $d\mu_{\xi}$ is a positive measure and $h$ a harmonic function on 
${\Bbb R}^{3}.$ (In (D) below, we briefly discuss the situation where 
(2.19) is not assumed). In particular, if $\nu $ is uniformly bounded 
above, say sup $\nu  =$ 0, then the Liouville theorem for harmonic 
functions implies that $h \equiv $ 0, and one has the expression
\begin{equation} \label{e2.21}
\nu (x) = -\int_{{\Bbb R}^{3}}\frac{1}{|x-\xi|}d\mu_{\xi}. 
\end{equation}
Note that since $\nu$ is harmonic on $\Omega$,
\begin{equation} \label{e2.22}
\Roof{I}{\bar} \subset  supp d\mu  \subseteq  \partial\Omega , 
\end{equation}
but in many situations, as will be seen below, the first inclusion is 
strict.

\medskip

{\bf (A)(i). Pure harmonic potentials.}

 Suppose that $\nu $ is a smooth harmonic function defined on all of 
${\Bbb R}^{3},$ so that $\nu  = h$ in (2.20). It is clear that in this 
case $\Sigma  = \emptyset  $ in the sense that $I = \emptyset  $ in 
${\Bbb R}^{3}.$ Since $\nu $ is axisymmetric, $\nu $ may be viewed as 
an expansion in Legendre polynomials, i.e.
$$\nu  = \sum_{k\geq 0}a_{k}R^{k}\cdot  P_{k}(\frac{z}{R}), $$where 
$R^{2} = r^{2}+z^{2}.$ For instance, $R\cdot  P_{1}(\frac{z}{R}) = z$, 
$R^{2}\cdot  P_{2}(\frac{z}{R}) = 3z^{2}- R^{2}.$

 While these solutions are defined on all of ${\Bbb R}^{3},$ no such 
solution gives a complete Weyl metric $g$ on $M = {\Bbb R}^{3},$ by 
Theorem 1.1. For instance, for $\nu  = z$, the Weyl metric is 
$$g = e^{-2z}(e^{-r^{2}}(dr^{2}+dz^{2}) + r^{2}d\phi^{2})),  u = e^{z}. 
$$
Any straight ray in the $(r,z)$ half-plane has finite length in this 
metric, except a ray parallel to the negative $z$-axis. The horizon 
$\Sigma $ occurs formally at $\{z = -\infty\},$ of infinite 
$g$-distance to any point in ${\Bbb R}^{3}$.

\medskip
{\bf (A)(ii). Newtonian potentials.}

 Suppose that $h = 0$ in (2.20), so that $\nu $ is the Newtonian 
potential of an axisymmetric positive mass distribution $d\mu $ as in 
(2.21). This situation corresponds exactly to the Newtonian theory of 
gravity, (or equivalently the electrostatics of a positively charged 
distribution). While there is a vast classical literature on this 
subject, we will only consider the most interesting situation where
\begin{equation} \label{e2.23}
supp d\mu  = \Roof{I}{\bar}, 
\end{equation}
so that $\nu $ approaches $-\infty $ on a dense set in supp $d\mu .$ 

 The following Lemma characterizes this situation.

\bbgin{lemma} \label{l 2.4.}
   Let $d\mu $ be an axi-symmetric positive Radon measure on ${\Bbb 
R}^{3}.$ Then 
\begin{equation} \label{e2.24}
supp d\mu  = \Roof{I}{\bar} \Leftrightarrow  supp d\mu  \subset  A. 
\end{equation}

\end{lemma}
{\bf Proof:}
 Suppose first that supp $d\mu $ is not contained in $A$. Since $d\mu $ 
is axially symmetric, part of supp $d\mu $, namely the part not 
contained in $A$, is then given by a union of circles about the 
$z$-axis. Suppose first that there is a circle $C$ which is an isolated 
component of supp $d\mu$, so that $d\mu |_{C}$ is a multiple of 
Lebesgue measure on $C$. This case has been examined in [Wl],[BW], and 
we refer there for details. In particular in this case the potential 
$\nu $ is bounded below on and near $C$, and hence supp $d\mu  \neq  
\Roof{I}{\bar}.$ If $C$ is not isolated, then using (2.21), the same 
reasoning holds, since the measure $d\mu$ is then even less 
concentrated on the circles.

 On the other hand, if supp $d\mu  \subset A$, then $d\mu $ is a 
positive Radon measure on $A$. Standard measure theory implies that the 
upper density of $d\mu $ w.r.t. Lebesgue measure $dA$ at $a\in A,$ i.e. 
$\limsup_{r\rightarrow 0}\frac{\mu (B_{a}(r))}{r},$ is positive, for 
Lebesgue almost all $a\in supp d\mu .$ From the expression (2.21), it 
is clear that for any such $a$, $\nu (x) \rightarrow  -\infty $ as $x 
\rightarrow  a$. This gives the converse.
{\endproof}

\medskip

 For the remainder of the discussion in (A), we assume (2.23) holds. 
From the theory of subharmonic functions on ${\Bbb R}^{3},$ the set $I$ 
given by (2.15) may be an arbitrary $G_{\delta}$ set in $A$ $\subset  
{\Bbb R}^{3},$ i.e. a polar set in $A$. Since countable unions of polar 
sets are polar, note that $I$ is not necessarily closed in 
$A\subset{\Bbb R}^{3}.$ (For example, let $\{z_{i}\} \in A$ be an 
increasing sequence converging to a limit point $z$, with say 
$dist(z_{i}, z_{i+1}) = i^{-2},$ and let $d\mu  = \sum 
2^{-i}\delta_{z_{i}},$ where $\delta_{z_{i}}$ is the Dirac measure 
based at $z_{i}.$ Then $I = \{z_{i}\}$ and supp $d\mu  = \{z_{i}\cup 
z\}).$

 Given any $x\in{\Bbb R}^{3},$ let $m_{x}(r)$ be the mass of the 
measure $d\mu $ in the ball $B_{x}(r),$ i.e.
\begin{equation} \label{e2.25}
m_{x}(r) = \int_{B_{x}(r)}d\mu . 
\end{equation}
This is a non-negative increasing function on ${\Bbb R}^{+},$ for any 
given $x$ and the limit
\begin{equation} \label{e2.26}
m =  lim_{r\rightarrow\infty}m_{x}(r) >  0, 
\end{equation}
is the total mass of $d\mu .$ This agrees, up to a universal constant 
factor, with the (ADM or Komar) mass in general relativity, when the 
latter is defined, and with (0.7) for solutions with pseudo-compact 
boundary. Note that one may have $m = +\infty .$ 

 Lemma 2.4 and a standard result from potential theory, c.f. [H, 
Thm.3.20], characterize the possible Riesz measures satisfying (2.23).

\bbgin{lemma} \label{l 2.5.}
  A necessary and sufficient condition that a positive Radon measure 
$d\mu $ is the Riesz measure of an axi-symmetric subharmonic function 
$\nu $ on ${\Bbb R}^{3}$ with sup $\nu  =$ 0 and supp $d\mu  = 
\Roof{I}{\bar}$ is that supp $d\mu  \subset $ A, and, for any given $x 
\in A$,
\begin{equation} \label{e2.27}
\int_{1}^{\infty}\frac{m_{x}(r)}{r^{2}}dr <  \infty. 
\end{equation}
\end{lemma}
{\endproof}

 It is easy to see that a Weyl solution $(M, g)$ generated by a 
potential $\nu $ as in (2.21) for which supp $d\mu  = \Roof{I}{\bar}$ 
is a {\it compact} subset of the axis $A$, is asymptotically flat, in 
the sense preceding Theorem 0.1. Further, the simplest or most natural 
surfaces enclosing any finite number of compact components of 
$\Roof{I}{\bar},$ and intersecting $A$ outside $\Roof{I}{\bar},$ are 
2-spheres in $M$. Of course if $suppd\mu\subset A$ is non-compact, then 
the solution cannot be asymptotically flat. A simple example is the 
solution generated by the measure
$$d\mu  = \frac{1}{1+|\zeta|}dA_{\zeta}, $$where $\zeta $ parametrizes 
$A$ and $dA$ is Lebesgue measure on $A$. Observe also that such 
solutions do not have pseudo-compact boundary.

\medskip

 It is worthwhile to discuss some standard examples of Weyl solutions 
and their corresponding measures.

\smallskip
\bbgin{example} \label{Example 2.6.}
{\bf (i).(Curzon Solution).}
 {\rm From the point of view of the Riesz measure, perhaps the simplest 
example is the measure $d\mu $ given by a multiple of the Dirac measure 
at some point on $A$, so that $\nu  = - m/R,$ $R(x) = |x|$, is a 
multiple of the Green's function on ${\Bbb R}^{3}.$ This gives rise to 
the Curzon (or monopole) solution, c.f. [Kr,(18.4)], 
\begin{equation} \label{e2.28}
g_{C} = e^{2m/R}[e^{- m^{2}r^{2}/R^{4}}(dr^{2} + dz^{2}) + 
r^{2}d\phi^{2}], 
\end{equation}
with $u = e^{- m/R}.$ Here $\Omega = {\Bbb R}^{3} \setminus \{0\}$, 
$\partial\Omega  =$ \{0\}, and it is often stated that $g_{C}$ has a 
point-like singularity (monopole) at the origin. However the geometry 
of $\partial M$ is very different than that of a point. Namely the 
circles about the $z$-axis have length diverging to infinity as $R 
\rightarrow $ 0. Thus, small spheres $R = \epsilon $ about \{0\} become 
very long in the $\phi $ direction, and very short in the transverse 
$\theta $ direction, forming a very long, thin cigar. In particular, as 
a metric space, $\partial M = {\Bbb R} .$ This is the first example 
where $\partial M$ is non-compact but pseudo-compact. Of course 
$\partial M = \Sigma ,$ so that $(M, g)$ is complete away from $\Sigma 
.$ A more detailed analysis of the Curzon singularity is given in [SS].

 Note that one could not have solutions with both directions expanding 
at $\partial M,$ so that area $\partial M = \infty ,$ with $\partial M$ 
pseudo-compact. This can be seen by use of minimal surface arguments, 
c.f. [G].}
\end{example}

{\bf (ii). (Schwarzschild Solution).}
 The Schwarzschild metric (0.4) is a Weyl metric, with measure
 $d\mu  = \frac{1}{2}dA$ on $[-m, m]$, where $dA$ is the standard 
Lebesgue measure on $A$. The resulting potential $\nu $ in (2.21) is 
the Newtonian potential of a rod on the z-axis with mass density 
$\frac{1}{2},$ given by

\begin{equation} \label{e2.29}
\nu_{S} = 
\frac{1}{2}
log ( \frac{{R_+} + {R_-} -  2m}{{R_+} + {R_-} + 2m} ),
 \ \ {\rm where} \ \  R_{\pm}^{2} = r^{2} + (z \pm  m)^{2}.  
\end{equation}

 As mentioned before, the horizon $\Sigma $ here is a smooth totally 
geodesic 2-sphere of radius $2m$ and $\partial M = \Sigma .$

 The Weyl solution generated by the potential $a\cdot \nu_{S},$ for 
$\nu_{S}$ as in (2.29) with $a$ $ > $ 0 and $a$ $\neq $ 1, is not 
isometric or homothetic to the Schwarzschild metric. The associated 
Weyl metric is no longer smooth up to the horizon; in fact $\Sigma $ is 
not even a 2-sphere unless $a = 1$.

\medskip
\bbgin{remark} \label{Remark 2.7.}
  {\rm More generally, consider any Weyl solution generated by a Riesz 
measure $d\mu $ satisfying (2.23). Observe that $f = \frac{r}{u},$ the 
length of the $\phi $ circles in the Weyl metric, stays bounded away 
from 0 and $\infty $ on approach to supp $d\mu ,$ if and only if
\begin{equation} \label{e2.30}
logr -  C \leq  \nu  \leq  logr + C, 
\end{equation}
for some $C <  \infty ,$ since $\nu = log u$. From the expression 
(2.21), this occurs only for the Schwarzschild potential $\nu_{S}.$ 
Briefly, the reason for this is as follows. The estimate (2.30) implies 
that the potential $\upsilon  = \nu-\nu_{S}$ is bounded and given by 
convolution of $dist^{-1}$ with a signed Radon measure $d\lambda .$ 
However, as in the proof of Lemma 2.4, if $\upsilon $ is bounded then 
one sees that necessarily $d\lambda  <<  dA$ and further the 
Radon-Nikodym derivative $d\lambda /dA$ is 0, a.e. $(dA)$. In other 
words, any point of non-zero density for $d\lambda$ w.r.t. $dA$ gives 
rise to approximating points on which $\nu$ is unbounded. It follows 
that $d\lambda  =$ 0, and hence $\upsilon  =$ 0.

 In all other cases, where $f \rightarrow $ 0 on approach to a region 
in supp $d\mu ,$ the length of the $\phi $ circles goes to 0, and hence 
this portion of $\Sigma $ is singular, of dimension $\leq $ 1, while 
where $f \rightarrow  \infty ,$ the length of the circles goes to 
$\infty $ and the corresponding portion of $\Sigma $ is singular and 
non-compact, (as in the Curzon solution). Note that if supp $d\mu $ is 
compact, then in all cases, $\partial M$ is pseudo-compact. 

 Thus among the Weyl solutions given by a Newtonian potential, only the 
Schwarzschild metric is smooth up to $\Sigma .$ This gives a very 
simple (local) confirmation of Theorem 0.1 in this special case.}
\end{remark}
\medskip

\noindent
{\bf Example 2.6.(iii). (Superposition/Multiple Holes).}
 Subharmonic functions of the form (2.21) form a convex cone. In 
particular, one thus has a natural linear superposition principle for 
Weyl solutions. This feature is another remarkable property of Weyl 
solutions. 

 For example, one may choose the measure $d\mu  = \frac{1}{2}dA$ on 
two, or any number of disjoint intervals $\{I_{j}\}$ on the axis $A$, 
provided (2.27) holds. These correspond to solutions with 'multiple 
black holes', each interval $I_{j}$ giving a component of $\Sigma $ 
which is a 2-sphere of radius equal to the length of $I_{j}.$ Although 
such solutions are essentially smooth up to $\Sigma ,$ they do not 
define smooth vacuum solutions on ${\Bbb R}^{3}\setminus \cup B_{j}.$ 
There are cone singularities, (called struts or rods in the physics 
literature), along geodesics (corresponding to $A\setminus \{I_{j}\})$ 
joining the 2-spheres of $\Sigma ,$ so that the metric $g$ is not 
locally Euclidean along such curves. Thus, the elementary flatness 
condition (2.14) is violated on $A\setminus \cup I_{j}.$ Nevertheless, 
the curvature of such metrics is uniformly bounded everywhere. These 
cone singularities must be considered part of $\partial M,$ so 
$\partial M$ in this case is a union of 2-spheres joined by a 
collection of curves and thus connected.

 Of course, the black hole uniqueness theorem, Theorem 0.1, also 
implies that such solutions cannot be smooth everywhere, when the 
number of intervals is finite. However, the proof of this result 
strongly uses the asymptotically flat assumption. It seems to be 
unknown whether there are any Schwarzschild type metrics with 
infinitely many black holes, i.e. metrics complete away from $\Sigma $ 
and smooth up to $\Sigma $ with $\Sigma $ consisting of infinitely many 
2-spheres, and which satisfy (2.17) or (2.21). It is natural to 
conjecture that such solutions do not exist, c.f. however the end of 
Remark 2.8 below. 

\medskip
\bbgin{remark} \label{Remark 2.8.}
 {\rm Whenever supp $d\mu  \subset  A$ is not connected but compact, 
there will exist such cone singularities on $A\setminus$ supp $d\mu$. 
When supp $d\mu ,$ or a sufficiently small smoothing of supp $d\mu ,$ 
is interpreted to represent a matter source, then this statement 
corresponds exactly to the fact that there are no equilibrium (i.e. 
time-independent) many-body solutions in Newtonian gravity of this 
character, c.f. [Bn]. Even when supp $d\mu$ is non-compact and 
disconnected, this seems very likely to be true, c.f. the expressions 
for sums of Schwarzschild rods (2.29) in [IK, p.336-337], which 
generalize to infinitely many rods.

 The components $\{C_{i}\}$ of $A\setminus$ supp $d\mu $ represent 
idealized matter sources (struts or rods) keeping the components of 
supp $d\mu $ apart in static equilibrium. The cone angle $\alpha  = 
\alpha_{i}$ is constant on each $C = C_{i},$ and corresponds to a 
concentration of scalar curvature on $C \subset  \Roof{M}{\bar}$ given 
by a multiple of the Lebesgue measure on $C$
$$s = (1-\alpha )dA_{C}, $$Thus the vacuum equations (0.1) are not 
satisfied across $\{C_{i}\}.$ If a very small tubular neighborhood of 
radius $r$ of such a rod is smoothly filled in with a perfect fluid 
source of constant pressure $\rho $ and energy density $\mu ,$ then one 
has the relation $lim_{r\rightarrow 0} r\rho  = -  lim_{r\rightarrow 0} 
r\mu  > $ 0. The effective mass of such rods is zero, i.e. they do not 
contribute to the gravitational potential $\nu ,$ c.f. [I2] for a 
detailed discussion.

 By passing to covering spaces, it is always possible to create such 
cone singularities in Weyl solutions, even if none existed to begin 
with. For instance, for the Schwarzschild solution (0.4), with 
potential (2.29), take any covering, including the universal covering, 
of ${\Bbb R}^{3}\setminus A = (S^{2}\setminus \{a\cup -a\})\times {\Bbb 
R}^{+},$ where \{$a$, $-a$\} are two antipodal points on $S^{2}.$ This 
gives a solution whose metric completion has cone singularities along 
two (radial) geodesics starting at the antipodal points on $S^{2} = 
\Sigma $ and going to infinity.

 Note that this discussion assumes that $\nu $ is given by a Newtonian 
potential (2.21). In fact, there are Weyl solutions $(M, g, u)$ 
everywhere smooth up to the axis $A$, with $\Sigma  = \partial M$ 
disconnected, with no cone singularities or struts keeping the 
components of $\Sigma $ apart. Namely, the B1 solution (0.9), dual to 
the Schwarzschild solution, has this property.

 Another, more remarkable, example is given in [KN]. These authors 
construct a Weyl solution of the form (2.12), which is complete away 
from $\Sigma$ and smooth up to $\Sigma$, with $\Sigma$ consisting of 
infinitely many Schwarzschild-like 2-spheres. In fact, the solution is 
periodic in the $z$-direction. This metric is not of the form (2.21), 
but is a limit of a sequence of solutions of the form (2.20), c.f. (D) 
below.}
\end{remark}

\bbgin{remark} \label{Remark 2.9.}
 {\rm If $(M, g, u)$ is a Weyl solution of the form (2.12) with $\nu  = 
\nu_{u} =$ $logu$, then the dual solution $(M' , g' , f)$, discussed in 
(2.9), is also a Weyl solution of the form (2.12), with potential 
$\nu_{f}$ = $logf$ given by
\begin{equation} \label{e2.31}
\nu_{f} = logr -  \nu_{u}. 
\end{equation}
Hence if one potential is Newtonian, the dual one is not. Note that the 
sets $I_{u}, I_{f}$ where $\nu_{u}$ and $\nu_{f}$ are $-\infty $ are 
disjoint, with $\Roof{I}{\bar}_{u}\cup\Roof{I}{\bar}_{f} = A$. Hence, 
if $\nu_{u}$ is a Newtonian potential with supp $d\mu $ compact, so 
that the associated Weyl solution is asymptotically flat, then the dual 
Weyl solution is asymptotically cylindrical.

 Another example of a potential where both terms in (2.20) are 
non-trivial is the situation considered (locally) in [GH], where $h$ is 
a smooth axi-symmetric harmonic function defined on a neighborhood of 
supp $d\mu ,$ c.f. Remark 1.5. As vacuum solutions, these metrics 
cannot be complete away from $\Sigma ,$ as in the discussion on pure 
harmonic potentials.}

\end{remark}

  This completes our discussion of the Positive Case.

\medskip

{\bf (B).(Negative Case).}
 Under the assumption (2.16), suppose now instead that $\nu $ is 
locally bounded below in ${\Bbb R}^{3},$ i.e. 
\begin{equation} \label{e2.32}
\inf_{B_{x}(1)}\nu  >  -\infty , \forall x\in\partial\Omega . 
\end{equation}
Then $\nu $ extends uniquely to a globally defined superharmonic 
function on
${\Bbb R}^{3}.$ Exactly the same discussion as in (A) above holds here,
under the substitution $\nu  \rightarrow  -\nu .$ (This corresponds to 
the
transformation $u \rightarrow  u^{-1}$ following (1.5)). In this case, 
the
Riesz measure is a negative measure, so that one has solutions with 
negative
mass. Note that here the potentials $\nu $ or $u$ are unbounded {\it 
above} within supp $d\mu ,$ i.e. $u$ or $\nu$ go to $+\infty$.

\medskip

{\bf (C).(Mixed Case).}
 Next, one may superimpose Weyl solutions with positive and negative 
measures $d\mu ,$ i.e. consider $\nu $ of the form (2.21), with $d\mu $ 
a signed Radon measure. For example, one may form dipole-type solutions 
with potential of the form $\nu  = \nu_{+} + \nu_{-}$ where $\nu_{+}$ 
and $\nu_{-}$ are (for instance) Curzon or Schwarzschild solutions of 
positive and negative mass placed at different regions on the axis. 
This gives for instance examples of asymptotically flat solutions with 
mass $m$ assuming any value in ${\Bbb R} .$

 More generally, since positivity is no longer assumed, the measure 
$d\mu $ may be replaced by distributions, for example weak derivatives 
of measures.

\bbgin{example} \label{Example 2.10.}
 {\bf (Multipole Solutions).}
 {\rm As a typical example, one may take potentials corresponding to 
derivatives of the Dirac measure based at a point $a\in A,$ i.e. the 
multipole potentials,
$$R^{-n-1}\cdot  P_{n}(\frac{z}{R}), $$where $P_{n}$ is the $n^{th}$ 
Legendre polynomial, or arbitrary linear combinations of such; c.f. 
[MF, p.1276ff].

 Such potentials are limits of combinations of Newtonian potentials 
with positive and negative mass. Thus, it is reasonable to expect that 
there are (Newtonian) equilibrium solutions, i.e. solutions with no 
cone singularities on the axis. This is proved in [Sz], where explicit 
equilibrium conditions are given. Note that one may have infinitely 
many multipole 'particles` in equilibrium.

 Another example is the potential of a dipole ring
$$\nu (x) = -\int_{C}\frac{z(x)}{|x-\xi|^{3}}d\xi , $$where $x = 
(r,z,\phi )$ and $d\xi $ is the Lebesgue measure on the unit circle $C 
= \{r=1\}$ in the $z =$ 0 plane. Here $\nu (x) \rightarrow  -\infty ,$ 
as $x \rightarrow  C$ along the rays $r =$ 1, $z > $ 0. Hence, in this 
case, the set $I = C$ is not contained in the axis $A$.}

\end{example}

{\bf (D). (Limits).}
 Finally, one may consider potentials which are limits of potentials of 
the type (A)-(C) above, (besides those in Remark 2.9, Example 2.10). We 
consider just one important instance of this here.

\bbgin{example} \label{Example 2.11.}
 {\bf (Kasner Metric).}
 {\rm It is easily seen that the potential $\nu  = logr$ generates the 
flat metric 
$$g = dr^{2}+dz^{2}+d\phi^{2}, $$on $({\Bbb R}^{3})^+$, with potential 
$u = r$. Observe that since the $\phi$-lines have constant length, the 
function $r$ is now an affine (in fact linear) function on $({\Bbb 
R}^{3})^+.$

 On the other hand, an equally simple computation shows that the 
potential $\nu  = a\cdot  logr,$ for any $a\in{\Bbb R} ,$ generates the 
metric
$$g = r^{2a^{2}-2a}(dr^{2}+ dz^{2}) + r^{2-2a}d\phi^{2},  u = r^{a}. 
$$Equivalently, setting $s = r^{a^{2}-a+1},$ 
\begin{equation} \label{e2.33}
g = ds^{2}+ s^{\alpha}dz^{2} + s^{\beta}d\phi^{2},  u = s^{\gamma}, 
\end{equation}
where $\alpha  = (2a-2)/(a-1+a^{-1}), \beta  = 
(2a^{-1}-2)/(a-1+a^{-1}), \gamma  = (a-1+a^{-1})^{-1}.$ Here $s\in{\Bbb 
R}^{+}, z\in{\Bbb R} $ and $\phi  \in  [0,2\pi ]$ or any other 
interval, including ${\Bbb R} ,$ (by passing to covering and quotient 
spaces). These metrics are all non-homothetic, provided $a\in 
[-1,0)\cup (0,1]$; $a = 0$ gives the flat metric with $u = 1$ while $a 
= -1$ gives the $A3$ metric (1.8). 

 The potential $\nu  = a\cdot  logr$ can be considered as the limit
\begin{equation} \label{e2.34}
\nu  = lim_{m\rightarrow\infty}a[\nu_{S}(m) -  log2m], 
\end{equation}
where $\nu_{S}(m)$ is the Schwarzschild potential (2.29) of mass $m$. 
Thus it is a limit of potentials of the form (2.18), where both terms 
are non-zero, (the harmonic term $h$ is of course constant here).

 These metrics are dual, in the sense discussed in (2.9) to the Kasner 
(or Bianchi I) vacuum cosmological models, with homogeneous (flat) but 
anisotropic space-like hypersurfaces, c.f. [Wd, Ch.7.2]. It is easy to 
see that the Kasner metrics are the only Weyl solutions $(M, g)$ which 
have an isometric ${\Bbb R}\times {\Bbb R} $ action, even locally. (The 
axisymmetric potential $\nu$ on $\Omega \subset {\Bbb R}^{3}$ must be 
invariant under an orthogonal ${\Bbb R}$-action, hence giving a 
rotationally invariant harmonic function on ${\Bbb R}^{2}$. Thus the 
potential must be a multiple of log $r$).

 Consider these metrics on the quotient $M = {\Bbb R}^{+}\times 
S^{1}\times S^{1}.$ In case $a > 0$, we have $\alpha  < 0$, $\beta  >  
0$, $\gamma  >  0$ and so $\Sigma  = \partial M.$ As in the discussion 
with the Curzon metric, the $z$-circles have unbounded length as $s 
\rightarrow $ 0, so that $\partial M = {\Bbb R} $ and the levels $t = 
\epsilon$, $(t(x) = dist_{g}(x, \partial M)),$ are long, thin cigars. 
Thus, $\partial M$ is non-compact, but pseudo-compact. The end of 
$\Roof{M}{\bar}\setminus U$ is obviously small.

 If $a < $ 0, then $\alpha  > $ 0, $\beta  > $ 0, $\gamma  < $ 0 and so 
$\Sigma  = \emptyset  ,$ (it occurs at infinity), with $\partial M =$ 
\{pt\}. In this case, the end of $\Roof{M}{\bar}\setminus U$ is {\it 
not} small; the area growth of geodesic spheres is $O(r^{1- \gamma})$.

 However, none of these solutions are asymptotically flat even in a 
weak sense, except of course when $a = 1$. Namely, the curvature decays 
only quadratically in the $g$-distance to $\partial M,$ i.e. 
$$|r| = O(t^{-2}), $$and not any faster. Hence, the case $a >$ 0 shows 
that the conclusion of Theorem 0.3, (asymptotically flat or small 
ends), cannot be strengthened to only asymptotically flat ends, while 
the case $a < $ 0 shows that the assumption (ii) on $u$ in Theorem 0.3 
is necessary.}

\end{example}

\medskip
 Another metric of this (limit) type is that constructed in [KN], 
referred to in Example 2.8. This metric has the same asymptotics as the 
Kasner metric.

\section{Characterization of Asymptotically Flat Solutions.}
\setcounter{equation}{0}

 In this section, we prove Theorem 0.3. The proof of the first 
statement on finiteness of the number of ends is quite easy, so we 
begin with this.

 Throughout this section, let $(M, g, u)$ be a static vacuum solution 
with $\partial M$ pseudo-compact. We recall from \S 0 that $M$ is connected
and oriented. As in \S 0, let
$$t(x) = dist_{g}(x, \partial M), $$
and suppose $U = t^{-1}(0, s_{o}),$ so that $\partial U \cap M$ is 
compact. For $r,s \geq  s_{o},$ let $S(s) = t^{-1}(s), A(r,s) = 
t^{-1}(r,s)$ be the geodesic spheres and annuli about $\partial U.$ It 
is important to note that neither $S(s)$ nor $A(r,s)$ are necessarily 
connected, even if $M$ has only one end. (Of course if $E$ is a given 
end of $M$, then $S_{E}(s) = S(s)\cap E$ must be connected for some 
sequence $s = s_{j} \rightarrow  \infty$). Let $S_{c}(s)$ and 
$A_{c}(r,s)$ denote any component of $S(s)$ resp. $A(r,s)$, so that 
$S(s) = \cup S_{c}(s), A(r,s) = \cup A_{c}(r,s).$ Of course $t$ is a 
proper exhaustion function on $\bar M\setminus U,$ so that these sets 
have compact closure in $M$. 

 Let $diam^{i}A_{c}(r,s)$ denote the intrinsic diameter of 
$A_{c}(r,s),$ i.e. the diameter of the connected metric space 
$(A_{c}(r,s), g)$.

\bbgin{lemma} \label{l 3.1.}
  There exists a constant $d_{o} <  \infty ,$ independent of $s$, such 
that the number of components of $A(\frac{1}{2}s,2s)$ is at most 
$d_{o}$ and
\begin{equation} \label{e3.1}
diam^{i}A_{c}(\tfrac{1}{2}s,2s) \leq  d_{o}\cdot  s. 
\end{equation}
In particular, the manifold $(\bar M\setminus U, g)$ has a finite 
number of ends $\{E_{i}\}.$
\end{lemma}
{\bf Proof:}
 Consider first the 4-manifold $(N, g_{N}),$ as in (0.2) which is 
smooth and Ricci-flat outside $\bar U = \pi^{-1}(U),$ where $\pi : N 
\rightarrow  M$ is projection on the first factor. Since $\partial\bar 
U\cap N$ is compact,  it follows from results of [Lu] that Lemma 3.1 
holds on $N$, so that in particular $N$ has a finite number of ends. 
Since $N = M\times S^{1}, \bar M\setminus U$ also has a finite number 
of ends.

 The choice of the time parameter on $N$ defines a totally geodesic 
embedding $M \subset  N$ and we have $t_{N}|_{M} = t$ where $t_{N}(x) = 
dist_{N}(x, \partial M), \partial M \subset  \Roof{N}{-}.$ A geodesic 
ball or annulus in $M$ embeds in the geodesic ball or annulus of the 
same size in $N$. Hence (3.1) also holds for $M$.

{\endproof}

 Lemma 3.1 of course proves the first statement of Theorem 0.3. Observe 
that the estimate (3.1) is invariant under rescaling of the metric $g$.

\medskip

 For the remainder of the proof, we (usually) work with a given end $E$ 
from the finite collection $\{E_{i}\}.$ The main statement of Theorem 
0.3 is that if
\begin{equation} \label{e3.2}
\int^{\infty}\frac{1}{area S_{E}(s)}ds <  \infty , 
\end{equation}
then the end $E$ is asymptotically flat. The proof of this result is 
rather long, so we outline here the overall strategy. The asymptotic 
behavior of $(E, g, u)$ is studied in general by examining the 
structure of the possible tangent cones at infinity, defined below. 
Basically, tangent cones at infinity fall into two classes, according 
to whether the asymptotic geometry near a given divergent sequence of 
base points is non-collapsing or collapsing, c.f. Lemmas 1.3-1.4. The 
main point is to prove that under the bound (3.2), all tangent cones at 
infinity are {\it flat}  manifolds, and further that no collapse 
behavior is possible. Once this is established, the proof that $(E, g, 
u)$ is asymptotically flat is relatively straightforward.

 Apriori, the end $E$ may be very complicated topologically, for 
instance of infinite topological type; consider for instance that $E$ 
might be of the form $S_{\infty}\times S^{1},$ where $S_{\infty}$ is 
any non-compact surface of infinite topological type and one end. A 
main idea is to use the behavior of the potential function $u$, in 
particular its value distribution theory, to control the topology and 
geometry of $E$ in the large. We have already seen in \S 2 that the 
potential $u$ controls quite strongly the geometry of Weyl solutions. 
Lemma 3.6 below is the key technical lemma which expresses this control 
for general static vacuum solutions (with pseudo-compact boundary). 
Further remarks on the strategy of proof precede the Lemmas below.

  The discussion to follow, until the end of Lemma 3.6, holds in 
general for ends $E$ of static vacuum solutions with compact boundary. 
The estimate (3.2) will only be used after this.

\medskip

 We now define the {\it  tangent cones at infinity}  of a given end 
$E$. (While this is a commonly used terminology, such limit metric 
spaces are not necessarily metric cones in general). 

 First, we recall by Theorem 1.1 that there is a constant $K <  \infty 
$ such that, $\forall x\in M,$
\begin{equation} \label{e3.3}
|r|(x) \leq  \frac{K}{t^{2}(x)}, |dlogu|(x) \leq  \frac{K}{t(x)}. 
\end{equation}
The scale-invariant estimates (3.3) give quite strong initial control 
on the asymptotic geometry of $(E, g, u)$ which allows one to get 
started. Observe that an immediate consequence of (3.1) and (3.3), by 
integration along paths in $A_{c}(\frac{1}{3}s,3s),$ is the following 
Harnack inequality:
\begin{equation} \label{e3.4}
\frac{sup u}{inf u} \leq  K_{1}, 
\end{equation}
where the sup and inf are taken over any component 
$A_{c}(\frac{1}{2}s,2s)$ and $K_{1}$ is independent of 
$A_{c}(\frac{1}{2}s,2s).$

 Let $x_{i}$ be any divergent sequence of points in $E$, with $t_{i} = 
t(x_{i}) \rightarrow  \infty .$ Consider the connected geodesic annuli 
$A_{i} = A_{i}(\kappa ) = A_{c}(\kappa^{-1}t_{i}, \kappa t_{i}), 
x_{i}\in A_{i},$ w.r.t. the rescaled or blow-down metric 
\begin{equation} \label{e3.5}
g_{i} = t_{i}^{-2}\cdot  g;
\end{equation}
here $\kappa $ is any fixed positive constant $> $ 1. By the curvature 
estimate (3.3), the metrics $g_{i}$ have uniformly bounded curvature on 
$A_{i}$ - the curvature bound depends only on $K$ and $\kappa .$ 
Further, by (3.1), the diameter of $A_{i}$ w.r.t. $g_{i}$ is also 
uniformly bounded. 

 Hence, if the sequence is non-collapsing, i.e. if there is a lower 
volume bound $vol_{g}A_{i} \geq  \nu_{o}\cdot  t_{i}^{3},$ for some 
$\nu_{o} > $ 0, (equivalent to $vol_{g_{i}}A_{i} \geq  \nu_{o}$ by 
scaling), then Lemma 1.3 implies that a subsequence of the pointed 
sequence $\{(A_{i}, g_{i}, x_{i})\},$ converges smoothly, away from the 
boundary, to a limiting smooth metric $(A_{\infty}(\kappa), g_{\infty}, 
x_{\infty}).$ The limit is a solution of the static vacuum equations; 
as noted in Lemma 1.3, the potential $u$ is renormalized to $u_{i} = 
u/u(x_{i}),$ so that the limit potential $u_{\infty}$ satisfies 
$u_{\infty}(x_{\infty}) =$ 1, c.f. also (3.4). Choosing a sequence 
$\kappa_{j} \rightarrow  \infty $ and a suitable diagonal subsequence, 
gives the maximal static vacuum solution $(A_{\infty}, g_{\infty}, 
u_{\infty}, x_{\infty}).$ Observe here also that the estimate (3.1) 
implies that $\partial A_{\infty} =$ \{pt\}.

 On the other hand, if the sequence $(A_{i}, g_{i})$ is collapsing, in 
the sense that $vol_{g}A_{i} <<  t_{i}^{3},$ as $t_{i} \rightarrow  
\infty ,$ (equivalent to $vol_{g_{i}}A_{i} \rightarrow $ 0), then Lemma 
1.4 implies that $A_{i}$ is a Seifert fibered space or torus bundle 
over an interval. As discussed there, one may then pass to ${\Bbb Z} $ 
or ${\Bbb Z}\oplus {\Bbb Z} $ covers $\widetilde A_{i}(\kappa)$ of 
$A_{i}(\kappa)$ (or more precisely smooth interior approximations to 
$A_{i}(\kappa))$ to obtain a non-collapsing sequence $(\widetilde 
A_{i}(\kappa), g_{i}, x_{i})$, smoothly convergent to a limit 
$(\widetilde A_{\infty}(\kappa), g_{\infty}, x_{\infty})$; ($g_{i}$ 
here is lifted to the cover $\widetilde A_{i},$ as is $x_{i}$). As 
above, one may then choose a sequence $\kappa_{j} \rightarrow  \infty $ 
and pass to a diagonal subsequence to obtain a maximal limit 
$(\widetilde A_{\infty}, g_{\infty}, u_{\infty}).$ This limit static 
vacuum solution has an isometric ${\Bbb R}$ action, or ${\Bbb R}\oplus 
{\Bbb R} $ action in the case of a rank 2 collapse.  Hence by 
Proposition 2.2 it is a Weyl solution. In the latter case, the solution 
is then a (possibly flat) Kasner metric, c.f. Example 2.11. Topologically,
the limit $\widetilde A_{\infty}$ is a trivial ${\Bbb R}$, 
(or ${\Bbb R}\oplus {\Bbb R}$), bundle over a surface $V$, (or interval),
again by Proposition 2.2.

 As in \S 2, we will always work in a ${\Bbb Z}$-quotient $\bar 
A_{\infty}$ of $\widetilde A_{\infty},$ (or ${\Bbb Z}\oplus{\Bbb Z} $ 
quotient in the case of rank 2 collapse), and finite covers $\bar 
A_{i}(\kappa)$ converging to $\bar A_{\infty}(\kappa).$ For any fixed 
$\kappa  > $ 0, the manifolds $(\bar A_{i}(\kappa ), g_{i}, x_{i})$ 
thus have uniformly bounded curvature and diameter, a uniform lower 
bound on their volume, and converge smoothly to the limit $(\bar 
A_{\infty}(\kappa ), g_{\infty}) \subset  (\bar A_{\infty}, 
g_{\infty}).$ The limit has a free isometric $S^{1}$ or $S^{1}\times 
S^{1}$ action, and so in particular is an $S^{1}$ or $T^{2}$ bundle. 
Hence $\bar A_{i}(\kappa)$ is also topologically an $S^{1}$ or $T^{2}$ 
bundle, (although not metrically).

  To be definite, the finite covers are chosen so that the length of 
the $S^{1}$ factor or factors at the base point $x_{i} \in \bar 
A_{i}(\kappa)$ converge to $1$ in the limit.

  Recall by Lemma 1.4 that the inclusion map of the fibers induces 
an injection on $\pi_{1}$. The coverings $\bar A_{i}(\kappa)$ are obtained 
by taking large finite unwrappings of the $S^{1}$ or $T^{2}$ fibers, 
(corresponding to taking subgroups of $\pi_{1}(S^{1})$ or 
$\pi_{1}(T^{2})$ of large but finite index). All finite covering spaces 
of $S^{1}$ or $T^{2}$ are 
still $S^{1}$ or $T^{2}$, and hence we may, and do, choose the unwrappings
so that, as smooth manifolds,
$$\bar A_{i}(\kappa ) = A_{i}(\kappa ), $$
for any $\kappa  > $ 0. In the limit, the 
unwrapping of the collapse thus just corresponds to expanding the 
length of the collapsing $S^{1}$ factor (or factors), preserving
the holonomy, if any, of the $S^{1}$ bundle; compare with the 
proof of Proposition 2.2. 

 The limit spaces $(A_{\infty}, g_{\infty}, u_{\infty}, x_{\infty})$ or 
$(\bar A_{\infty}, g_{\infty}, u_{\infty}, x_{\infty})$ constructed 
above are called tangent cones at infinity of $(E, g, u)$. Note that 
such tangent cones are only attached to some subsequence of a given 
divergent sequence of base points $\{x_{i}\}.$ Hence, apriori, the 
tangent cones at infinity could be highly non-unique as Riemannian 
manifolds. In general, there may be no relation between the geometry of 
different tangent cones based on (subsequences of) distinct divergent 
sequences $\{x_{i}\};$ for example, tangent cones based on sequences 
with $t(x_{i}) = 2^{i^{2}}$ and $t(x_{i}) = 2^{i^{3}}.$ The tangent 
cones only detect behavior of the end $E$ in $g_{i}$-bounded distance 
to the base points $x_{i}.$

  On the other hand, since tangent cones at infinity attached to any 
divergent sequence always exist, for any $s$ sufficiently large, say $s 
\geq s_{o}$, the geometry of $(A_{c}(\frac{1}{2}, 2), g_{s})$ or $(\bar 
A_{c}(\frac{1}{2}, 2), g_{s})$ is always close to that of {\it some} 
tangent cone at infinity $A_{\infty}$ or $\bar A_{\infty}$. Further, by
construction, the tangent cones are always connected and, since $M$ is
oriented, so is each tangent cone.

\medskip

 The following lemma is a typical application of the use of tangent 
cones at infinity.
\bbgin{lemma} \label{l 3.2.}
  Suppose the curvature $r$ decays faster than quadratically in the end 
$(E, g, u)$, i.e.
\begin{equation} \label{e3.6}
|r|(x) \leq  \frac{\varepsilon (t)}{t^{2}(x)}, 
\end{equation}
where $\varepsilon (t) \rightarrow $ 0 as $t \rightarrow  \infty .$ 
Then there is a compact set $K \subset  E$ such that $E \setminus K$ is 
diffeomorphic either to ${\Bbb R}^{3} \setminus B$ or to $({\Bbb 
R}^{2}\setminus B)\times S^{1},$ where $B$ is a 3-ball, (resp. a 
2-ball), i.e. $E$ is of standard topological type. Further, the annuli 
$A_{E}(\frac{1}{2}s, 2s)$ are connected, for all $s \geq s_{o}$, for 
some $s_{o} < \infty$.
\end{lemma}
{\bf Proof:}
  By the preceding discussion, the condition (3.6) is equivalent to the 
statement that all tangent cones at infinity $(A_{\infty}, g_{\infty})$ 
or $(\bar A_{\infty}, g_{\infty})$ of $E$ are {\it flat}, (as well as
connected and oriented). The two 
possible conclusions of Lemma 3.2 correspond to the two possibilities 
of non-collapse and collapse in the formation of the tangent cones. 

 Suppose first that $g$ is non-collapsing on $E$, i.e. there exists 
$\nu_{o} > $ 0 such that $vol_{g}A_{c}(\frac{1}{2}s,2s) \geq  
\nu_{o}\cdot  s^{3},$ for all $s$ large, and all components $A_{c}$. 
Recall that $\partial A_{\infty} = \{pt\}$ in this situation. It then 
follows that for $s$ sufficiently large, each $A_{c}(\frac{1}{2}s,2s)$ 
is diffeomorphic, and almost isometric to the standard flat annulus $A 
= r^{-1}(\frac{1}{2}s,2s)$ in ${\Bbb R}^{3},$ $r(x) = |x|,$ (away from 
the boundary). In fact, each tangent cone at infinity $A_{\infty}$ is 
isometric to ${\Bbb R}^{3} \setminus \{0\}$ in this situation. Here we
are implicitly using the fact that the only complete oriented flat 
3-manifold with an isolated singularity is ${\Bbb R}^{3} \setminus \{0\}$,
c.f. [AC] for example. Similarly, a smooth approximation to $S_{c}(s)$ 
is diffeomorphic and almost isometric to $S^{2}(s)\subset{\Bbb R}^{3}.$ 

  By the isotopy extension theorem, these diffeomorphisms from 
$A_{c}(\frac{1}{2}s,2s)$ to the standard annulus may then be assembled 
to a global diffeomorphism, and almost isometry, of $E \setminus K$ 
into ${\Bbb R}^{3}\setminus B,$ for some compact set $K \subset E$. 
We refer to 
[AC, Thm.1.18] for the proof of these statements, (in a slightly 
different but equivalent form), which are now quite standard. The main 
point is of course that since the family of annuli 
$A_{c}(\frac{1}{2}s,2s)$ as $s$ varies is topologically rigid, i.e. one 
has a unique topological type, there is no value of $s$ at which the 
topology can change or bifurcate.

 Suppose on the other hand that $g$ is collapsing on 
$A_{c}(\frac{1}{2}s_{i},2s_{i}),$ for some sequence $s_{i} \rightarrow  
\infty $, and some sequence of components $A_{c}$. As discussed above, 
one may then pass to suitable covers $\Roof{A}{\bar}_{i} = 
\Roof{A}{\bar}_{c}(\frac{1}{2}s_{i},2s_{i})$ so that, in a subsequence, 
$(\Roof{A}{\bar}_{i}, g_{i})$ is diffeomorphic and almost isometric to 
its limit $\bar A_{\infty}(\frac{1}{2}, 2) \subset \bar A_{\infty}$. 
The maximal limit $\bar A_{\infty}$ is a flat manifold with either a 
free isometric $S^{1}$ or $S^{1} \times S^{1}$ action. Hence there are 
two possibilities for $\bar A_{\infty}$, namely either $V \times S^{1}$ 
or ${\Bbb R}^{+} \times S^{1} \times S^{1}$, where $V$ is a flat 
2-manifold and the metric is a product metric on each $S^{1}$ factor. 
In the former case, the diameter estimate (3.1) implies, (as in the 
non-collapse case above), that $V$ is a complete flat cone, possibly 
with an isolated singularity at $\{0\}$. Hence, although these two 
possibilities for the limiting metric of $\bar A_{\infty}(\frac{1}{2}, 
2)$ are distinct, both are the same topologically, i.e. $\bar 
A_{\infty}(\frac{1}{2}, 2)$ is topologically $I \times S^{1} \times 
S^{1}$.

  Now recall from the discussion on tangent cones that 
$\Roof{A}{\bar}_{i}$ is {\it diffeomorphic} to 
$A_{c}(\frac{1}{2}s_{i},2s_{i})$; metrically $\bar A_{i}$ approximates 
one of the types of flat manifolds above, with $S^{1}$ factors 
shrinking to very short circles. In both cases, (a smoothing of) 
$\Roof{S}{\bar}_{c}(s_{i})$ is diffeomorphic and almost isometric to a 
flat torus $T^{2}.$ 

  In particular, the topological type of $\bar A_{i}$ is distinct from 
that of the annuli $A_{i}$ above in the non-collapse case, which are 
topologically always of the form $ I \times S^{2} \subset {\Bbb R}^{3} 
\setminus \{0\}$, (for any choice of base point and component). This 
implies first that the family $\{A_{c}(\frac{1}{2}s,2s)\}$ must be 
collapsing for all $s$, as $s \rightarrow  \infty$, and all components 
$A_{c}$. Second, the topological type of the annuli $\bar A_{s}$, and 
hence that of $A_{s} = A_{c}(\frac{1}{2}s, 2s)$ is unique, and given 
for all $s$ large and all $c$ by $I \times S^{1} \times S^{1}$. Use of 
the isotopy extension theorem in the same way as above then proves that 
the end $E$ itself is diffeomorphic to $({\Bbb R}^{2} \setminus B) 
\times S^{1}$.

{\endproof}

\begin{remark} \label{r 3.3.}{\bf (i)}
{\rm It is easily seen from the vacuum equations (0.1) that the 
condition (3.6) is equivalent to
\begin{equation} \label{e3.7}
|\nabla logu|(x) \leq  \frac{\varepsilon (t)}{t(x)}, 
\end{equation}
as $t \rightarrow  \infty ,$ c.f. also the proof of Theorem 1.1.} 

{\bf (ii).}
  {\rm In the context of Lemma 3.2, the tangent cones at infinity 
$(\bar A_{\infty}, g_{\infty})$ may not be unique up to isometry in the 
collapse case, and so may vary within the moduli space ${\cal M}_{1}$ 
of flat product metrics of the form $V \times S^{1}$ or within the 
moduli space ${\cal M}_{2}$ of flat product metrics of the form ${\Bbb 
R}^{+} \times S^{1} \times S^{1}$. 

  Note that the moduli space ${\cal M}_{o}$ of flat metrics on ${\Bbb 
R}^{3}$ or ${\Bbb R}^{3} \setminus \{0\}$ is just one point, (c.f. 
again [AC] for the latter statement for example). Similarly, by the 
normalization preceding Lemma 3.2 that the $S^{1}$ factors have length 
1, the moduli space ${\cal M}_{2}'\subset {\cal M}_{2}$ normalized in 
this way is also just one point. The moduli space ${\cal M}_{1}' 
\subset {\cal M}_{1}$ where the $S^{1}$ factor has length 1 is 
naturally identified with ${\Bbb R}^{+}$, parametrized by the cone 
angle at $\{0\}$.

  Observe however that these two moduli spaces ${\cal M}_{1}'$ and 
${\cal M}_{2}'$ are disjoint; they cannot be connected (or even 
approximated) by a curve of flat metrics. Now the geometry of the 
annuli $(A_{E}(\frac{1}{2}s, 2s), g_{s})$ varies continously with $s$. 
By the remarks preceding Lemma 3.2, this induces a continuous variation 
of the possible tangent cones $(\bar A_{\infty}, g_{\infty})$ in ${\cal 
M}_{1}'$ or ${\cal M}_{2}'$. Hence, on a given end $E$, one cannot 
obtain two different tangent cones, one of the form $V \times S^{1}$ 
and another of the form ${\Bbb R}^{+} \times S^{1} \times S^{1}$; c.f. 
also the proof of Lemma 3.7 below.}

{\bf (iii).}
  {\rm We will need a slight generalization of Lemma 3.2 for the next 
lemma below. Thus let $\gamma(s)$ be any properly embedded curve in $E$ 
with $t(\gamma(s)) \rightarrow \infty$ as $s \rightarrow \infty$, and 
suppose (3.6), (or (3.7)), holds in the balls $B_{\gamma(s)}(\delta 
\cdot t(\gamma(s)))$, for some fixed $\delta > 0$. Then the conclusion 
of Lemma 3.2 also holds.

  To see this, consider the blow-downs $g_{s} = t^{-2}(\gamma(s)) \cdot 
g$, based at $\gamma(s)$, and the associated tangent cones at infinity. 
The scale invariant condition (3.6), when applied to 
$B_{\gamma(s)}(\delta \cdot t(\gamma(s)))$, implies that all such 
tangent cones are flat in $(B_{x_{\infty}}(\delta), g_{\infty})$, and 
thus flat everywhere in their maximal domain $A_{\infty}$, by the fact 
that smooth solutions of the vacuum equations are real analytic. The 
proof then proceeds exactly as in Lemma 3.2}
\end{remark}

  To prove that the estimates (3.6) and (3.7) do in fact hold on $E$, 
we need to understand in more detail the value distribution of the 
potential $u$. The main result needed for this is given in Lemma 3.6, 
and then (3.6)-(3.7) follow rather easily in (3.17)-(3.18) below. 
However, some preliminary results are required for the proof of Lemma 
3.6. The main difficulty is that $u$ may not, in this generality, be a 
proper function onto its image, (c.f. the remark following the proof of 
Lemma 3.4). Lemma 3.4 below is a slightly weaker substitute for this 
property.

\medskip

 Let $U = t^{-1}(0, s_{o})$ be a neighborhood of $\partial M$ as in the 
beginning of \S3. Let $\gamma (\tau )$ be a maximal flow line of 
$\nabla u.$ We will say that $\gamma $ is {\it  divergent}  if $\gamma 
$ does not intersect $U$ at two different times, i.e. if $\gamma (\tau 
)$ exits $U$ at some time, then $\gamma (\tau )$ never reenters (a 
possibly distinct component of) $U$, and if further $\gamma (\tau )$ 
does not terminate at a critical set of $u$ in $\Roof{M}{\bar}\setminus 
U$ as $\tau  \rightarrow  \pm\infty .$ It follows that if $\gamma $ is 
divergent, then $\gamma $ is complete in at least one direction, $(\tau 
 \rightarrow  +\infty , \tau  \rightarrow  -\infty $ or both), and in 
any such direction, $\gamma (\tau )$ diverges to infinity in $M$. Since 
the potential $u$ has no local maxima or minima in $M$, the set of flow 
lines terminating on a critical set of $u$ in $M\setminus U$ is a closed
set of measure 0 in $M$. This follows for instance from the fact that the
measure $|du|dA$, where $dA$ is Lebesgue measure on the level sets of $u$,
is preserved under the gradient flow of $u$, and this measure tends to
0 on approach to critical points of $u$.
Thus among the set of flow lines not joining points 
of $U$, the divergent flow lines are generic in terms of measure on 
$M$. Of course the flow lines are curves of steepest ascent for $u$ as 
$\tau$ increases, and of steepest descent for $u$ as $\tau$ decreases.

\bbgin{lemma} \label{l 3.4.}
   There exists a compact set $K \subset  {\bar M}$, with ${\bar U} 
\subset K$, such that any divergent flow line $\gamma (\tau )$ 
intersects K.

\end{lemma}
{\bf Proof:}
 Suppose that this were not the case, so that there exist, necessarily 
complete, flow lines $\gamma (\tau ), \tau\in{\Bbb R} ,$ which do not 
intersect a given $K \supset  {\bar U}$. We may choose $K$ sufficiently 
large so that $\gamma (\tau )$ is then contained in a fixed end $E 
\subset  \Roof{M}{\bar}\setminus U,$ since there are only finitely many 
ends. 

  Let $A_{\gamma}(\tau) = A_{\gamma}(\frac{1}{2}t(\gamma(\tau)), 
2t(\gamma(\tau)))$ be the component of the geodesic annulus containing 
the base point $\gamma(\tau)$ and let $E_{\gamma}$ be the part of $E$ 
swept out by such annuli, $E_{\gamma} = \cup_{\tau}A_{\gamma}(\tau) 
\subset E$. As $\tau \rightarrow \infty$, the function 
$u(\gamma(\tau))$ is monotone increasing. 

  If $u(\gamma(\tau))$ increases to $+\infty$, then $u \rightarrow 
\infty$ uniformly as $\tau \rightarrow \infty$ in $E_{\gamma}^{+} = 
\cup_{\tau > 0}A_{\gamma}(\tau) \subset E$, by the Harnack estimate 
(3.4). Since $E$ is an end, there exists some sequence $\tau_{j} 
\rightarrow \infty$ such that, for $t_{j} = t(\gamma(\tau_{j}))$, the 
spheres $S_{\gamma}(t_{j}) \subset A_{\gamma}(\tau_{j})$ satisfy 
$S_{\gamma}(t_{j}) = S_{E}(t_{j})$, i.e. the spheres $S_{E}(t_{j})$ are 
connected, and hence $A_{\gamma}(\tau_{j}) = A_{E}(\tau_{j})$. Thus, 
$u$ becomes uniformly unbounded on $A_{E}(\tau_{j})$, as $j \rightarrow 
\infty$. However, as $\tau \rightarrow -\infty$, the curve 
$\gamma(\tau)$ also diverges to infinity in $E$ and $u(\gamma(\tau))$ 
is decreasing, (and so in particular bounded), as $\tau \rightarrow 
-\infty$. This contradiction implies that $u(\gamma(\tau)) \rightarrow 
u^{+} < +\infty$, as $\tau \rightarrow +\infty$. Of course $u^{+} > 0$.

  Suppose first, (for simplicity), that $limsup_{E}u = u^{+}$, i.e. 
$lim_{t \rightarrow \infty}m(t) = u^{+}$, where $m(t) = sup_{S_{E}(t)}u$.
The maximum principle for the harmonic function $u$ implies that for
$t$ sufficiently large, the function $m(t)$ is either monotone increasing
or monotone decreasing in $t$ and hence approaches the value $u^{+}$ as
$t \rightarrow \infty$.
Consider the annuli $A_{\gamma}(\tau)$ in the scale $g_{t} = t^{-2} 
\cdot g$, $t = t(\gamma(\tau))$, as in (3.5). Any sequence $\tau_{i} 
\rightarrow \infty$ has a subsequence such that the corresponding 
annuli $(A_{\gamma}(\tau_{i}), g_{t_{i}})$ converge to a limiting 
domain $A_{\infty}(\frac{1}{2}, 2)$ in a tangent cone at infinity 
$(A_{\infty}, g_{\infty})$, or $(\bar A_{\infty}, g_{\infty})$, passing 
to covers as described above in the case of collapse. By construction, 
we then see that the potential function $u_{\infty}$ for this limit 
static vacuum solution achieves its maximal value $u^{+} > 0$ at an 
interior point. Since $u_{\infty}$ is harmonic, the maximum priniple 
implies that $u_{\infty} \equiv u^{+}$, and hence by the vacuum 
equations (0.1), the limit $(A_{\infty}, g_{\infty})$ or $(\bar 
A_{\infty}, g_{\infty})$ is flat. This argument holds for any 
subsequence, and since the convergence to the limit is smooth, we see 
that 
$$t^{2}\cdot  r|_{A_{\gamma}(\frac{1}{2}t,2t)} \rightarrow  0, \ \ t = 
t(\gamma(\tau)), \ \ {\rm as} \ \  \tau \rightarrow  +\infty, $$
by the scale-invariance of this expression.

 It follows from Lemma 3.2 and Remark 3.3(iii) that the end $E$ is 
topologically standard, and the annuli $A_{E}(\frac{1}{2}t,2t)$, $t = 
t(\gamma(\tau))$, are connected in $E$, for all $\tau$ sufficiently 
large. From the prior argument, this implies in particular that $u 
\rightarrow u^{+}$ uniformly at infinity in $E$.

  However, as before, as $\tau \rightarrow -\infty$, $u(\gamma(\tau))$ 
is monotone decreasing to a value $u_{-} \geq 0$. It follows that 
$u^{+} = u_{-}.$ This is of course impossible, and shows that $\gamma 
(\tau )$ must have exited $E$ at some (negative) time.

 Thus it remains to prove that $L \equiv limsup_{E} u = u^{+}$. Since the
annuli $A_{E}(\tau_{j})$ above are connected, the Harnack inequality
(3.4), together with the fact that $u^{+} < \infty$, implies that 
$L < \infty$. Now choose points $x_{j} \in S_{\gamma}(t_{j})$ such that
$u(x_{j}) \rightarrow L$. As above, the functions $u|_{A_{E}(\tau_{j})}$
have a subsequence converging to a limit harmonic function $u_{\infty}$ 
on a tangent cone at infinity based at $x_{\infty} = lim x_{j}$. 
Then as before $u$ has an interior maximum at $x_{\infty}$ and hence 
$u_{\infty} \equiv L$; this gives $L = u^{+}$.

{\endproof}

\medskip
 It follows from Lemma 3.4 and the discussion preceding it that any 
maximal flow line of $\nabla u$ intersects an apriori given large 
compact set $K \subset  M$, except those exceptional flow lines which 
start or end at a critical point of $u$ far out in $M$. In particular, 
a set of full measure in any given level set $L$ of $u$ may be 
connected to points in $K$ by flow lines of $\nabla u.$ In this sense, 
$u$ is 'almost proper`, in that it behaves almost like a proper 
function in terms of the gradient flow.

 Observe that this does not necessarily imply that the level sets of 
$u$ are compact, i.e. that $u$ is proper. For instance, the Weyl 
solution generated by the dipole potential $\nu  = \nu_{+} + \nu_{-}$ 
considered in \S 2(IIC) satisfies (3.6), (it is even asymptotically 
flat), but the 0-level of $\nu $ is non-compact if $\nu_{+}$ and 
$\nu_{-}$ are chosen so that the mass is 0. In this example, the only 
divergent flow lines of $\nabla u$ are the two ends of the $z$-axis.

\medskip

 Next, as in \S 2, let
$$\nu  = logu .$$

  The following result is quite standard.
\bbgin{lemma} \label{l 3.5.}
   On $(N, g_{N})$ as in (0.2), with Riemannian metric, we have
$$\Delta_{N}|\nabla\nu| \geq  0. $$
\end{lemma}
{\bf Proof:}
 This standard estimate is a simple consequence of the 
Bochner-Lichnerowicz formula
$$\tfrac{1}{2}\Delta|\nabla\nu|^{2} = |D^{2}\nu|^{2} + <\nabla\Delta\nu 
, \nabla\nu>  + r(\nabla\nu , \nabla\nu ), $$on $(N, g_{N})$, where we 
have dropped the subscript $N$ from the notation. Since $\Delta\nu  =$ 
0, by (1.6), and since $(N, g_{N})$ is Ricci-flat, this gives
$$
\tfrac{1}{2}\Delta|\nabla\nu|^{2} \geq  |D^{2}\nu|^{2}. $$One computes
$$\Delta|\nabla\nu| =  
\tfrac{1}{2}|\nabla\nu|^{-1}\Delta|\nabla\nu|^{2} -  
\tfrac{1}{4}|\nabla\nu|^{-3}|\nabla|\nabla\nu|^{2}|^{2}, $$
and, by the Cauchy-Schwarz inequality $|\nabla|\nabla\nu|^{2}|^{2} \leq 
 4|D^{2}\nu|^{2}|\nabla\nu|^{2},$ so the result follows.

{\endproof}

  Lemmas 3.4 and 3.5 lead to the following key result relating the 
behavior of $|\nabla u|$ to the area growth of geodesic spheres. This 
result is a straightforward consequence of the divergence theorem for 
proper harmonic functions $u$ on manifolds of non-negative Ricci 
curvature. Lemma 3.4 allows one to remove the assumption that $u$ is 
proper.

\bbgin{lemma} \label{l 3.6.}
  There is a constant $C <  \infty $ such that for any component 
$S_{c}(s)$ of $S(s) \subset M$, $ s \geq 1$,
\begin{equation} \label{e3.8}
sup_{S_{c}(s)}|\nabla u| \leq  C\cdot  areaS_{c}(s)^{-1}. 
\end{equation}

\end{lemma}
{\bf Proof.}
 We work on the Riemannian 4-manifold $(N, g_{N})$ until the end of the 
proof. Let $\Roof{A}{\hat}_{c} = \Roof{A}{\hat}_{c}(s)   
= \pi^{-1}(A_{c}(\frac{1}{2}s,2s))$ and 
$\Roof{S}{\hat}_{c} = \Roof{S}{\hat}_{c}(s) = 
\pi^{-1}(S_{c}(s)),$ where $\pi : N \rightarrow  
M$ is projection on the first factor, with $S^1$ fibers. From the 
coarea formula, we have
\begin{equation} \label{e3.9}
\int_{\Roof{A}{\hat}_{c}}|\nabla\nu|^{2} = \int_{v}   
\int_{L_{v}\cap\Roof{A}{\hat}_{c}}|\nabla\nu|d\sigma_{v}dv, 
\end{equation}
where $L_{v}$ is the $v$-level set of $\nu $ in $N$ and the outer 
integral in (3.9) is over the range of values in ${\Bbb R} $ of $\nu $ 
in $\Roof{A}{\hat}_{c}.$ Now as remarked following Lemma 3.4, up to a 
set $\Roof{Z}{\hat}_{v}$ of measure 0 in $L_{v}\cap\Roof{A}{\hat}_{c},$ 
all points in the set $(L_{v}\cap\Roof{A}{\hat}_{c})\setminus 
\Roof{Z}{\hat}_{v}$ may be joined by flow lines of $\nabla\nu$ to 
points in a fixed bounded hypersurface $\Roof{T}{\hat}$ in 
$\Roof{K}{\hat}$, independent of $v$, $s$; here $K$ is the compact set 
from Lemma 3.4, and $\Roof{T}{\hat} = \partial \Roof{K}{\hat}$ for 
instance. Hence, by the divergence theorem applied to the harmonic 
function $\nu $ on $N$,
\begin{equation} \label{e3.10}
\int_{L_{v}\cap\Roof{A}{\hat}_{c}}|\nabla\nu| \leq  
\int_{\Roof{T}{\hat}}|\nabla\nu| \leq  c_{1}, 
\end{equation}
for some $c_{1} <  \infty,$ independent of $s$ and 
$\Roof{A}{\hat}_{c}$. 

 Now $|\nabla\nu|$ is subharmonic on $(N,
g_{N})$ by Lemma 3.5, and 
$diam_{N}^{i}\Roof{A}{\hat}_{c} \leq  c\cdot  s,$ by (the proof of) 
Lemma 3.1. A standard sub-mean value inequality for manifolds of 
non-negative Ricci curvature, c.f. [SY,Thm.II.6.2], then gives
\begin{equation} \label{e3.11}
sup_{\Roof{S}{\hat}_{c}(s)}|\nabla\nu|^{2} \leq  \frac{c_{2}}{vol 
\Roof{A}{\hat}_{c}}\int_{\Roof{A}{\hat}_{c}}|\nabla\nu|^{2}. 
\end{equation}
Hence the estimates (3.9)-(3.11) imply
\begin{equation} \label{e3.12}
sup_{\Roof{S}{\hat}_{c}(s)}|\nabla\nu|^{2} \leq  c_{3}\cdot  
osc_{\Roof{A}{\hat}_{c}}\nu\cdot  (vol\Roof{A}{\hat}_{c})^{-1}. 
\end{equation}
To estimate the right hand side of (3.12), again by (the proof of) 
Lemma 3.1, osc $\nu  \leq  c_{4}\cdot  sup|\nabla\nu|s$ on 
$\Roof{A}{\hat}_{c}.$ Further, we claim that
$$sup_{\Roof{A}{\hat}_{c}(s)}|\nabla \nu| \leq c 
sup_{\Roof{S}{\hat}_{c}(s)}|\nabla \nu|,$$ 
for some constant $c$ independent of $s$ and $\Roof{A}{\hat}_{c}$. 
To see this, by scale-invariance, it suffices to prove that 
$sup_{\Roof{A}{\hat}_{c}(1)}|\nabla \nu| \leq c 
sup_{\Roof{S}{\hat}_{c}(1)}|\nabla \nu|$ w.r.t. the rescaled metrics
$g_s = s^{-2}g$. By the curvature and diameter bounds (3.3) and (3.4)
and Lemmas 1.3 and 1.4, the metrics $(\Roof{A}{\hat}_{c} (1), g_{s})$
form a compact family of metrics in the $C^{\infty}$ topology,
unwrapping in the case of collapse. Thus, one has uniform control on
the metrics $g_{s}$ on $\Roof{A}{\hat}_{c}(1)$.
Similarly, when normalized if necessary by additive and multiplicative
constants so that $sup_{\Roof{S}{\hat}_{c}(1)} \nu =
sup_{\Roof{S}{\hat}_{c}(1)}|\nabla \nu| = 1$, the positive harmonic
functions $\nu$ on $(\Roof{A}{\hat}_{c}(1), g_{s})$ also form a compact
family of functions in the $C^{\infty}$ topology, i.e. a normal family. 
This follows by the Harnack
estimate (3.4) and the Harnack principle (elliptic regularity) for harmonic 
functions, c.f. [GT, Thm 2.11, Ch. 8]. This compactness of the metrics 
and functions from elliptic theory proves the claim above.

  Similarly, the metric compactness above also implies there is a constant
$c < \infty$ such that
$$c^{-1} \cdot area S_{c}(1) \leq  volA_{c}(1) \leq  c \cdot areaS_{c}(1),$$
w.r.t. the metrics $g_{s}$. (This estimate can also be derived directly
from the Bishop-Gromov volume comparison theorem). Rescaling back
to the metric $g$ then gives
\begin{equation} \label{e3.13}
c_{5}^{-1}\cdot  s\cdot  area S_{c}(s) \leq  volA_{c} \leq  c_{5}\cdot  
s\cdot  areaS_{c}(s), 
\end{equation}
for some constant $c_{5} <  \infty .$ Note that by definition,
\begin{equation} \label{e3.14}
area\Roof{S}{\hat}_{c}(s) = \int_{S_{c}(s)}udA, 
\end{equation}
and the same for $vol\Roof{A}{\hat}_{c}.$ Hence by (3.4), the estimate 
(3.13) holds also for $\Roof{S}{\hat}_{c}(s)$ and $\Roof{A}{\hat}_{c}$ 
in place of $S_{c}(s)$ and $A_{c}$. 

 Thus, by combining these estimates above, (3.12) gives
$$sup_{\Roof{S}{\hat}_{c}(s)}|\nabla\nu| \leq  c_{6}\cdot  
area\Roof{S}{\hat}_{c}(s)^{-1}. $$Using (3.14) and (3.4) again, this 
estimate implies (3.8).

{\endproof}

\medskip

  We are now in a position to begin the proof of Theorem 0.3 itself. 
Observe that the previous results in \S 3 have not used the assumption 
(3.2), nor the assumption (ii) in Theorem 0.3 that $u$ does not 
approach 0 everywhere at infinity in $E$. Only the assumption that 
$\partial M$ is pseudo-compact has been used. Hence, at this stage, we 
do not even know that $E$ has finite topological type. The main point 
initially is to prove that the estimates (3.6)-(3.7) above do hold on 
$E$ under these assumptions.

\medskip

  Recall that we have $S_{E}(s) = \cup S_{c}(s),$ for $S_{c}(s) \subset 
E$. Each geodesic ray $\sigma (s)$ in $E$, i.e. an integral curve of 
$\nabla t,$ with $\sigma (s)\in S_{E}(s),$ determines a component 
$S_{\sigma}(s) = S_{c}(s)$ s.t. $\sigma (s)\in S_{c}(s);$ the union of 
such components sweep out a part $E_{\sigma}$ of the end $E$. Of course 
$E$ is the union of $E_{\sigma}$ among all (non-homotopic) rays $\sigma 
.$ 

 From Lemma 3.1 and from the obvious $areaS_{E}(s) = \sum 
areaS_{c}(s),$ we have
$$\int^{\infty}areaS_{E}(s)^{-1}ds < \infty  \Leftrightarrow  
\int^{\infty}areaS_{\sigma}(s)^{-1}ds < \infty ,$$ 
for some geodesic ray $\sigma\subset E.$ (Here the integrals start at 
some fixed value $s \geq  s_{o} > $ 0).

  Hence, under the assumption (3.2), we have
\begin{equation} \label{e3.15}
\int^{\infty} area S_{\sigma}(s)^{-1}ds  \leq K < \infty,
\end{equation}
for some ray $\sigma \subset E$ and constant $K$.

 By integrating along the curve $\sigma ,$ (3.15), Lemma 3.6 and the 
Harnack estimate (3.4) imply that $u$ is uniformly bounded in 
$E_{\sigma}.$ In fact, we claim that
\begin{equation} \label{e3.16}
u_{\infty} = lim_{t(x)\rightarrow\infty}u(x) <  \infty  
\end{equation}
exists, where the limit is taken in $E_{\sigma}.$ To see this, let 
$\gamma (s)$ be any "quasi-geodesic" in $E_{\sigma},$ i.e. $\gamma $ is 
a smooth curve with $\gamma (s)\in S_{\sigma}(s)$ and $|d\gamma /ds| 
\leq  C_{1},$ for some $C_{1} <  \infty .$ By (3.8) and (3.15), we then 
have $\int^{\infty}|du(\gamma (s))|ds \leq  C\cdot  C_{1}\cdot  K <  
\infty ,$ and so $u(\gamma (s_{1})) -  u(\gamma (s_{2})) \rightarrow $ 
0 whenever $s_{1}, s_{2} \rightarrow  \infty .$ Hence the limit 
$u_{\infty}(\gamma )$ is well-defined. The diameter estimate (3.1) 
implies that all points in $E_{\sigma}$ lie on quasi-geodesics, (with a 
fixed $C_{1}$), in $E_{\sigma},$ starting on $S_{\sigma}(s_{o}).$ 
Further, the limit $u_{\infty}(\gamma )$ is clearly independent of 
$\gamma ,$ since for instance (3.8) and (3.15) imply that 
$$osc_{A_{\sigma}(\frac{1}{2}s_{j}, 2s_{j})} u \rightarrow  0, $$
on some sequence $s_{j} \rightarrow  \infty .$ Hence (3.16) follows.

 Next, since $E$ is an end, there exists some sequence $t_{j} 
\rightarrow  \infty $ such that the geodesic spheres $S_{E}(t_{j})$ are 
connected, and hence $S_{\sigma}(t_{j}) = S_{E}(t_{j}).$ By (3.16), 
$u|_{S_{E}(t_{j})} \rightarrow  u_{\infty}$ as $t_{j} \rightarrow  
\infty .$ The maximum principle applied to the harmonic function $u$ 
thus implies that $u|_{A_{E}(t_{j}, t_{k})} \rightarrow  u_{\infty},$ 
whenever $t_{j}, t_{k} \rightarrow  \infty .$ Thus, we see that (3.16) 
holds where the limit is taken in the {\it full} end $E$, and not just 
in $E_{\sigma}.$

 Now we use the assumption (ii) of Theorem 0.3, which says that 
$u(x_{j}) \geq  u_{o} > $ 0, for some constant $u_{o} > $ 0 and some 
divergent sequence $x_{j} \in  E$. It follows from this and the 
existence of the limit (3.16) in $E$ that 
$$u_{\infty} > 0. $$
Hence, we may, and will, renormalize the potential function $u$ of the 
static vacuum solution $(M, g, u)$ so that, on $E$,
\begin{equation} \label{e3.17}
lim_{t(x)\rightarrow\infty}u(x) = 1. 
\end{equation}

 The estimate (3.17) essentially immediately implies the 
scale-invariant estimates
\begin{equation} \label{e3.18}
  sup_{S_{E}(s)}|r| <<  s^{-2}, \  sup_{S_{E}(s)}|\nabla u| <<  s^{-1}, 
\ {\rm as} \ s \rightarrow  \infty , 
\end{equation}
strenthening the bounds (3.3). For as discussed following (3.3), (3.18) 
is equivalent to the statement that all tangent cones at infinity of 
$E$ are flat, (c.f. also the proof of Lemma 3.2). But, as noted above, 
all tangent cones at infinity are static vacuum solutions, and (3.17) 
implies that the limit potential $u_{\infty}$ satisfies $u_{\infty} 
\equiv $ 1. Hence, the static vacuum equations (0.1) of course imply 
the limit metrics are flat.

 Lemma 3.2 now determines the topology of the end $E$, as one of two 
(standard) alternatives, according to non-collapse or collapse behavior 
at infinity.

 Before proceeding to the analysis of these cases, note that (3.18) 
implies that the metrics $g$ and $\Roof{g}{\tilde} = u^{2}\cdot  g$ 
from (1.4) are quasi-isometric on $E$, and almost isometric near 
infinity. Since $\Roof{g}{\tilde}$ has non-negative Ricci curvature on 
$E$, standard volume comparison theory implies that the area and volume 
ratios
\begin{equation} \label{e3.19}
\frac{area_{\Roof{g}{\tilde}}(\Roof{S}{\tilde}(s))}{s^{2}} , 
\frac{vol_{\Roof{g}{\tilde}}(\Roof{B}{\tilde}(s))}{s^{3}} 
\end{equation}
are monotone non-increasing in $s$. Hence, their limits at $s = \infty 
$ exist, and by the equality of $g$ and $\Roof{g}{\tilde}$ at infinity, 
the limits at $s = \infty $ of the ratios in (3.19) w.r.t. the $g$ 
metric and $g$-geodesic spheres and balls also exist, and equal the 
$\Roof{g}{\tilde}$ limits.

\medskip

 To proceed further, we now separate the discussion into non-collapse 
and collapse cases.

\medskip

{\bf Case A. (Non-Collapse).}

  Suppose that $E$ is non-collapsing at infinity, i.e. by the remarks 
above,
\begin{equation} \label{e3.20}
limsup_{s\rightarrow\infty}\frac{v(s)}{s^{3}}  =  
lim_{s\rightarrow\infty}\frac{v(s)}{s^{3}}   >  0, 
\end{equation}
where $v(s)$ denotes $vol_{g}(B_{E}(s)).$ This implies, via (3.13), 
that $areaS(s)^{-1} \leq  c\cdot  s^{-2},$ and hence, by Lemma 3.6, 
\begin{equation} \label{e3.21}
sup_{S(s)}|\nabla u| \leq  cs^{-2}. 
\end{equation}

 The volume condition (3.20) implies that all tangent cones at infinity 
$A_{\infty}$ of $(E, g)$ exist, without passing to covering spaces, 
and, as discussed above, are flat solutions of the static vacuum 
equations with potential $u_{\infty} \equiv 1$. As discussed in the 
proof of Lemma 3.2, the tangent cone at infinity is unique, (up to 
isometry), and given by ${\Bbb R}^{3} \setminus \{0\}$, and further $E 
\setminus K$ is diffeomorphic to ${\Bbb R}^{3} \setminus B$, for some 
compact set $K \subset E$. The blow-down metrics $g_{s} = s^{-2}\cdot 
g$ on all annuli $A_{E}(\frac{1}{2}s, 2s)$ converge smoothly to the 
flat metric on $A(1,2) \subset {\Bbb R}^{3}$, uniformly as $s 
\rightarrow \infty$, and hence there are local (harmonic) coordinates 
on $A_{E}(\frac{1}{2}s, 2s)$ in which $g$ has the expansion $g_{ij} = 
\delta_{ij} + \gamma_{ij}$, where $|\gamma_{ij}(x)| \rightarrow $ 0 
uniformly as $s \rightarrow  \infty $. Again as discussed in the proof 
of Lemma 3.2, these local coordinates, e.g. on $\{A_{E}(2^{i-1}, 
2^{i+1})\}$, $i > 0$, may be assembled into a global chart, mapping $E 
\setminus K$ onto ${\Bbb R}^{3} \setminus B$, c.f. [AC] for further 
details if desired. With respect to such a chart, the metric $g_{ij}$ 
has the form
\begin{equation} \label{e3.22}
g_{ij} = \delta_{ij} + \gamma_{ij}, 
\end{equation}
on all of $E\setminus K$, with $|\gamma_{ij}| \rightarrow 0$ uniformly 
at infinity in $E$. In other words, $g$ is $C^{o}$ asymptotic to the 
flat metric at infinity.

 To prove that the metric on $E$ is (strongly) asymptotically flat, as 
defined preceding Theorem 0.1, consider again the metric 
$\Roof{g}{\tilde} = u^{2}\cdot  g.$ From (1.4) and (3.21), the 
curvature of $\Roof{g}{\tilde}$ decays as
$$|\Roof{r}{\tilde}| \leq  Ct^{-4},$$ as $t \rightarrow \infty .$
(Note also that $t$ and $\Roof{t}{\tilde}$ are approximately equal for 
$t$ large). 

 It follows that the expansion (3.22) may be improved, for 
$\Roof{g}{\tilde},$ to
\begin{equation} \label{e3.23}
\Roof{g}{\tilde}_{ij} = \delta_{ij} + O(t^{-2}), 
\end{equation}
in a suitable (harmonic) coordinate chart. We refer to [BKN] or [BM] 
for instance for further details here. Briefly, elliptic regularity 
theory applied to the equations (1.4)-(1.5), together with the 
curvature decay above, implies that the $2^{nd}$ derivatives of the 
metric $\Roof{g}{\tilde}$ in the coordinate chart decay as $t^{-4},$ so 
that the metric $\Roof{g}{\tilde}$ decays to the flat metric at a rate 
of $t^{-2}.$ Hence
\begin{equation} \label{e3.24}
g_{ij} = u^{-2} \Roof{g}{\tilde}_{ij} = (1+2\upsilon )\delta_{ij} + 
O(t^{-2}), 
\end{equation}
where $\upsilon  = 1- u.$ Here we are using that fact that since 
$|\nabla u| = O(t^{-2}),$ $u$ has an expansion of the form $u =$ 1 $+ 
O(t^{-1}).$ Further, since log $u$ is harmonic w.r.t. 
$\Roof{g}{\tilde},$ the decay (3.21) and (3.23) implies that 
$\Delta_{f}logu = O(t^{-4})$ for $t$ large, where $\Delta_{f}$ is the 
flat Laplacian on ${\Bbb R}^{3}.$ This means that $u$ has an expansion 
$u =$ 1 $-  \frac{m}{t} + O(t^{-2}),$ where $m$ is the mass of $E$ 
defined in (0.7).

 In particular, these estimates show that the end $E$ is asymptotically 
flat in the sense preceding Theorem 0.1. 

  Note that, to first order in $t^{-1},$ the function $\upsilon = 1 - 
u$ corresponds to the Green's function in ${\Bbb R}^{3},$ i.e. the 
fundamental solution of the Laplacian, weighted by the mass $m$. It is 
of course possible to have $m =$ 0, as for instance for the dipole-type 
Weyl solutions in \S 2(IIC), or also $m < 0$. Further, since $u$ has 
been normalized so that $u \rightarrow 1$ at infinity in $E$, the 
expression (0.7) for the mass is equivalent to the usual definition
\begin{equation} \label{e3.25}
m_{E} = \frac{1}{4\pi}\int_{S_{E}(s)}<\nabla u, \nabla t> dA,
\end{equation}
where $s$ is sufficiently large so that $S_{E}(s) \cap \partial E = 
\emptyset$. This is because the expression (3.25) is independent of 
$s$, since $u$ is harmonic, and the fact that it is asymptotic to the 
expression (0.7) as $s \rightarrow \infty$.

 This completes the analysis of Case A.

\medskip

{\bf Case B. (Collapse).}

 Under the standing assumption (3.15), suppose that the end $E$ is 
collapsing at infinity, i.e.
\begin{equation} \label{e3.26}
limsup_{s\rightarrow\infty}\frac{v(s)}{s^{3}}  =  
lim_{s\rightarrow\infty}\frac{v(s)}{s^{3}}   = 0.
\end{equation}
  
  We will prove that this situation is impossible. The results 
preceding Case A remain valid, so that (3.17)-(3.18) hold, all tangent 
cones at infinity $\bar A_{\infty}$ of $E$ are flat products of the 
form $V \times S^{1}$ or ${\Bbb R}^{+} \times S^{1} \times S^{1}$, 
where $V$ is a flat 2-dimensional cone. Further $E \setminus K$ is 
diffeomorphic to $({\Bbb R}^{2} \setminus B) \times S^{1}$, for some 
compact set $K \subset E$.

 By (3.15) and Lemma 3.6, we have
\begin{equation} \label{e3.27}
\int^{\infty} sup_{S(s)} |\nabla u|(s)ds \leq K_{1} <  \infty . 
\end{equation}

 The main point is now to show that $(E, g)$ itself, (and not just its 
tangent cones), is asymptotic to a flat quotient of ${\Bbb R}^{3},$ and 
hence has at most quadratic volume growth of geodesic balls or linear 
area growth of geodesic spheres. This is done in the following result, 
which is a strengthening of Lemma 3.2.
\bbgin{lemma} \label{l 3.7.}
  Under the assumptions (3.26) and (3.27) above, there is a compact set 
$K \subset  E$ such that $(E \setminus K, g)$ is quasi-isometric to a 
flat product $({\Bbb R}^{2} \setminus B)\times S^{1}$ or ${\Bbb 
R}^{+}\times S^{1}\times S^{1}.$
\end{lemma}
{\bf Proof:}
 As in Case A, it is useful to work with the metric $\tilde g = 
u^{2}g;$ again, this makes no significant difference, since (3.17) 
holds. {\it All} the metric quantities below are thus in the $\tilde g$ 
metric. For notational simplicity however, we drop the tilde from the 
notation.

 Let $t_{\infty}(x) = lim_{s\rightarrow\infty}(dist(x, S_{E}(s)) -  
s)$. As in the construction of Busemann functions, the limit here 
exists, c.f. [Wu] for a discussion of such functions. By construction, 
$t_{\infty}$ is a Lipschitz distance function, i.e. $t_{\infty}$ 
realizes everywhere the distance between its level sets. Observe that 
on ${\Bbb R}^{3}, t_{\infty}$ is just the distance function to 
$\{0\}\in{\Bbb R}^{3},$ on $V\times S^{1}, t_{\infty}$ is the distance 
function to $\{0\}\in V$ pulled back to $V\times S^{1},$ for any cone 
$V$ with vertex \{0\}, while on ${\Bbb R}^{+}\times S^{1}\times S^{1}, 
t_{\infty}$ is the distance function on ${\Bbb R}^{+}$ pulled back to 
the total space.

 By renormalization, (as with the potential $u$), $t_{\infty}$ induces 
a distance function $\bar t_{\infty}$ on each tangent cone $\bar 
A_{\infty}$ by defining $\bar t_{\infty}(x) = 
lim(t_{\infty}(x)/t_{\infty}(x_{i})),$ where $t(x_{i}) \rightarrow  
\infty $ and $x_{i}$ are the base points converging to the base point 
$x_{\infty} \in \bar A_{\infty}.$ Thus $\bar t_{\infty}$ is the 
function above on $V\times S^{1}$ or ${\Bbb R}^{+}\times S^{1}\times 
S^{1}.$

 The map $t_{\infty}: (E, \tilde g) \rightarrow  {\Bbb R}^{+}$ is 
distance non-increasing, and preserves distance along the integral 
curves of $t_{\infty};$ thus where smooth, it is a Riemannian 
submersion. We will show that $t_{\infty}$ gives rise to a Lipschitz 
quasi-isometry by examining the asympotics of the second fundamental 
form of its level sets.

 Thus, let $\sigma (s)$ be any geodesic ray in $E$ which is an integral 
curve of $\nabla t_{\infty},$ and let $B = B_{\sigma}(s)$ denote the 
second fundamental form of the level surface $t_{\infty}^{~-1}(s)$ at 
$\sigma (s).$ The form $B$ is well-defined and smooth along any such 
ray $\sigma$. Recall that $B$ satisfies the Riccati equation
\begin{equation} \label{e3.28}
B'  + B^{2} + R_{T} = 0, 
\end{equation}
where $T$ is the unit tangent vector along $\sigma .$ Consider the 
behavior of $s\cdot  B(s)$ as $s \rightarrow  \infty .$ This quantity 
is scale-invariant, and thus converges smoothly, (in subsequences), to 
the limiting expression $\bar s \cdot  B_{\infty}(\bar s)$ on any 
tangent cone at infinity. Since the parameters $s$ and 
$t_{\infty}|_{\sigma}$ are the same up to additive constants, $\bar s = 
\bar t_{\infty}.$ Similarly, by the definition of $\bar t_{\infty}, 
B_{\infty}$ is the second fundamental form of the levels $\bar 
t_{\infty}.$ Hence, either 

\noindent
(i): $\bar s\cdot  B_{\infty}(\bar s) = (d\theta /|d\theta|)^{2},$ when 
the tangent cone is of the form $V\times S^{1},$ and $\theta $ is the 
angle variable about $\{0\}\in V,$ or 

\noindent
(ii):  $\bar s\cdot  B_{\infty}(\bar s) =$ 0, when the tangent cone is 
of the form ${\Bbb R}^{+}\times S^{1}\times S^{1}.$ 

  Thus $\bar s\cdot  B_{\infty}(\bar s)$ is either of rank 1, with 
eigenvalue 1, or identically 0. Note that the expression in case (i) is 
independent of the cone $V$, i.e. the cone angle at $\{0\}$.

 As noted preceding Lemma 3.2 and in Remark 3.3(ii), the geometry of 
$\bar A(\frac{1}{2}s, 2s)$ smoothly approximates that of a limit 
tangent cone $\bar A_{\infty},$ for $s$ large, and varies continuously 
in $s$. Since the two alternatives (i) and (ii) above for the structure 
of $B_{\infty}$ are rigid, it follows that all tangent cones $\bar 
A_{\infty}$ are of the same type, i.e. they are all of the form 
$V\times S^{1},$ or all of the form ${\Bbb R}^{+} \times S^{1} \times 
S^{1}.$

 The main task now is to show that the deviation of $s\cdot  B(s)$ from 
its limit $\bar s\cdot  B_{\infty}(\bar s)$ has bounded integral. To do 
this, we use the Riccati equation, and estimate the decay of the 
curvature term $R_{T},$ using basically standard methods in comparison
geometry, c.f. [P, Ch.6.2] for instance.

 Thus, from (1.4), the sectional curvature $K$ of $(M, \tilde g)$ 
satisfies
$$K_{XZ} = |\nabla\nu|^{2} \geq  0,  \ \ K_{XY} = -|\nabla\nu|^{2}\leq  
0, $$
where $Z = \nabla u/|\nabla u|,$ $X,Y$ are vectors orthogonal to $Z$, 
and $\nu  =$ log $u$. Hence $|R_{T}| \leq  |\nabla\nu|^{2}.$ 
Substituting this in (3.28) gives
\begin{equation} \label{e3.29}
|B'  + B^{2}| \leq  |\nabla\nu|^{2}. 
\end{equation}
Let $\lambda $ be any eigenvalue of $B$, with unit eigenvector $e$; 
(note that $B$ is symmetric). Observe then that $s\cdot \lambda (s)$ 
converges either to 1 or to 0, as $s \rightarrow  \infty .$ The 
estimate (3.29) when applied to $(e, e)$ gives
$$\pm (s\lambda'  + s\lambda^{2}) \leq  s|\nabla\nu|^{2}. $$
Integrate this by parts along any finite interval $I$ to obtain
$$|s\lambda|_{\partial I}  + \int_{I} (s\lambda^{2} -  \lambda )ds| \leq  
\int_{I} s|\nabla\nu|^{2}ds. $$
If $s \lambda \rightarrow 0$ as $s \rightarrow \infty$, choose $I$ to be
any interval on which $s \lambda = 0$ at $\partial I$ and $s \lambda \neq 0$ 
on $I$, so that $s \lambda$ has a definite sign on $I$.
If there are no such boundary points, choose $I$ to 
be an infinite half-line. Similarly, if $s \lambda \rightarrow 1$, choose
$I$ to be intervals such that $s \lambda = 1$ at $\partial I$ with
$s \lambda - 1 \neq 0$ on $I$. Then summing up the estimate above over
all such intervals gives
$$\int_{\sigma} |s\lambda^{2} -  \lambda |ds \leq  
\int_{\sigma} s|\nabla\nu|^{2}ds + C_{o}, $$
for some constant $C_{o} < \infty$. Now the estimate (3.18), together 
with (3.27) gives
$$\int s|\nabla\nu|^{2}ds \leq  C_{1}, $$
for some constant $C_{1} <  \infty .$

 Suppose first $s\cdot \lambda (s) \rightarrow $ 1 as $s \rightarrow  
\infty .$ We then obtain
$$\int |s\lambda (\lambda  -  \frac{1}{s})|ds \leq  C_{2} <  \infty , $$
and hence
\begin{equation} \label{e3.30}
\int |\lambda  -  \frac{1}{s}|ds \leq  C_{3} <  \infty . 
\end{equation}
Similarly, if $s\cdot \lambda (s) \rightarrow $ 0 as $s \rightarrow  
\infty ,$ one obtains
\begin{equation} \label{e3.31}
\int|\lambda| ds \leq  C_{3} <  \infty . 
\end{equation}

 Now the second fundamental form $B$ gives the logarithmic derivative 
of the norm of Jacobi fields formed by the family of $t_{\infty}$-rays 
in $E$ starting at some level $t_{\infty}^{-1}(s_{o}).$ Thus, if $J$ is 
any Jacobi field formed from the $t_{\infty}$-conguence, and $v = 
J/|J|,$ we have along any $t_{\infty}$-ray,
$$B(v,v) = \frac{d}{ds}(log|J|(s)). $$
Suppose first that the end $E$ has tangent cones at infinity of the 
form ${\Bbb R}^{+} \times S^{1} \times S^{1} .$ Then (3.31) implies the 
uniform bound
\begin{equation} \label{e3.32}
C_{4}^{-1} \leq  |J|(s) \leq  C_{4}, 
\end{equation}
with $C_{4} = e^{C_{3}}.$ This means that the geometry of the level 
surfaces of $t_{\infty}$ is uniformly bounded as $s \rightarrow  \infty 
,$ i.e. the diameter and area of the level surfaces is uniformly 
bounded away from 0 and $\infty .$ It is then clear that there is a 
Lipschitz quasi-isometry of $(E\setminus K, \tilde g)$ to ${\Bbb R}^{+} 
\times S^{1} \times S^{1} $ induced by $t_{\infty}.$ Since $\tilde g$ 
and $g$ are also quasi-isometric, by (3.17), the lemma follows in this 
case.

 If $E$ has tangent cones at infinity of the form $V \times S^{1} ,$ 
then there is a basis of Jacobi fields whose elements satisfy either 
(3.32), or, from (3.30),
\begin{equation} \label{e3.33}
C_{4}^{-1}s \leq  |J|(s) \leq  C_{4}s. 
\end{equation}
As before, this implies that $t_{\infty}$ gives rise to a 
quasi-isometry of $(E\setminus K, g)$ to $({\Bbb R}^{2}\setminus B) 
\times S^{1}.$

{\endproof}

 Of course Lemma 3.7, in both cases, immediately implies that
\begin{equation} \label{e3.34}
areaS_{E}(s) \leq  c\cdot  s, 
\end{equation}
for some constant $c <  \infty .$ However, (3.34) violates the standing 
assumption (3.15), (or (3.2)). It follows that no end $(E, g)$ can 
satisfy the assumptions of Case B. 

  Together with Case A, this completes the proof of the second 
statement in Theorem 0.3.

\medskip

  We now turn to the proof of the last statement in Theorem 0.3. We 
will assume that the end $E$ is small, i.e.
\begin{equation} \label{e3.35}
\int_{E}\frac{1}{area S_{E}(s)} ds = \infty,
\end{equation}
and derive a contradiction from the assumptions $sup u < \infty$ and 
$m_{E} \neq 0$. 

  The proof is based on a result of Varopoulos [V] which states that 
ends of Riemannian manifolds satisfying (3.35) are {\it parabolic}, 
i.e. admit no non-constant positive superharmonic functions $v$ which 
tend uniformly toward their infimum at infinity. (Actually, the result 
in [V] is a condition on the volume growth of geodesic balls, but this 
is equivalent to the bound (3.35) under the estimate (3.13)). We will 
prove that the potential $u$ is such a non-constant function, giving 
the required contradiction.

  Thus, suppose first that
\begin{equation} \label{e3.36}
sup_{E} u <  \infty , 
\end{equation}
Arguing as in (3.9), but now on $E \subset M$ in place of $N$, we have
\begin{equation} \label{e3.37}
\int_{E}|\nabla u|^{2} = \int_{v}  \int_{L_{v}\cap E}|\nabla 
u|dA_{v}dv, 
\end{equation}
and as in (3.10),
$$\int_{L_{v}\cap E}|\nabla u| \leq  \int_{T}|\nabla u| \leq  C. 
$$Thus, these estimates imply that
\begin{equation} \label{e3.38}
\int_{E}|\nabla u|^{2} <  \infty ,
\end{equation}
so that
\begin{equation} \label{e3.39}
\int_{A_{E}(s,\infty )}|\nabla u|^{2} \rightarrow  0 \ \  {\rm as} \ \ 
s \rightarrow  \infty . 
\end{equation}

 On the other hand, again referring to the proof of Lemma 3.6, since 
almost all (in terms of measure) points in $L_{v}\cap E$ may be joined 
by flow lines of $\nabla u$ to a fixed bounded surface $T$ in $K$, by 
the divergence theorem there is a subsurface $T'  \subset  T$ such that
\begin{equation} \label{e3.40}
\int_{L_{v}\cap E}|\nabla u| \geq  \int_{T'}|\nabla u| \geq  c, 
\end{equation}
for some constant $c > $ 0. It follows from (3.37)-(3.40) that
\begin{equation} \label{e3.41}
osc_{A_{E}(s,\infty )}u \rightarrow  0 \ \ {\rm as} \ \ s \rightarrow 
\infty .
\end{equation}
By assumption (ii) in Theorem 0.3, we may thus assume w.l.o.g. that
\begin{equation} \label{e3.42}
lim_{t\rightarrow\infty}u = 1,
\end{equation}
in $E$. Thus, as noted in (3.18), $|r| \leq  \varepsilon (t)/t^{2},$ 
where $\varepsilon (t) \rightarrow $ 0 as $t \rightarrow  \infty$, and 
so Lemma 3.2 holds on $E$.

  Now choose a smooth approximation $S$ to a large geodesic sphere 
$S_{E}(s)$ with $S_{E}(s) \cap \partial E = \emptyset$, so that $S$ 
separates $E$ into two components, one being the outside containing the 
end $E$.
  
 We claim that if, in addition to (3.36),
\begin{equation} \label{e3.43}
m_{E} \neq 0,
\end{equation}
then there is a set of flow lines $\gamma$ of $\nabla u$ or $-\nabla 
u$, starting on $S$, of positive measure on $S$, and pointing out of 
$S$, which never intersects $S$ again at later times. Hence such 
$\gamma$ diverge to infinity in $E$, since up to a set of measure 0, 
$\gamma$ does not terminate in a critical point of $u$. (Compare with 
the earlier discussion regarding divergent flow lines and Weyl dipole 
solutions concerning Lemma 3.4).

  To see this, suppose instead that all flow lines say of $\nabla u$ 
which initially point out of $S$ eventually intersect $S$ again, with 
the exception of those terminating in critical points. Consider the 
measure
\begin{equation} \label{e3.44}
d\mu = <\nabla u, \nu>dA,
\end{equation}
on $S$, where $dA$ is the Lebesgue measure and $\nu$ is the unit 
outward normal on $S$; $d\mu$ is absolutely continuous w.r.t. $dA$. 
Since $u$ is harmonic, the divergence theorem implies that the gradient 
flow of $u$ preserves the measure $d\mu$ in the following sense. Let 
$D$ be a domain in $S$ and let $\Omega$ be the domain in $E$ formed by 
a collection of flow lines outside $S$, whose endpoints form a smooth 
surface $D'$. If $\nu'$ denotes the unit outward normal to $\Omega$ at 
$D'$, then the flow from $D$ to $D'$ carries the measure $d\mu$ to the 
measure $d\mu' = <\nabla u, \nu'>dA'$, where $dA'$ is Lebesgue measure 
on $D'$. In particular, the flow preserves the masses of the measures.

 Thus, under the assumptions above, the gradient flow (with varying 
flow-times), induces a homeomorphism of $S \setminus Z$ into itself, 
where $Z$ is a set of Lebesgue measure 0, corresponding to flow lines 
terminating in critical points. However, this homeomorphism inverts the 
direction of $\nabla u$ w.r.t. the fixed normal $\nu$, and hence maps 
domains $D_{+}$ on which the measure $d\mu$ is positive onto domains 
$D_{-}$ on which $d\mu$ is negative, in such a way that that 
\begin{equation} \label{e3.45}
m_{\mu}(D_{+}) = |m_{\mu}(D_{-})|.
\end{equation}
This of course implies that the total mass of $d\mu$ on $S$ is 0. 
However using (3.44) and the remarks concerning (3.25), the mass 
$m_{E}$ of $E$ equals the total mass of $d\mu$. This contradiction 
proves the claim.

 We may now complete the proof as follows. Assuming $E$ is an end 
satisfying (3.36) and (3.43), there exists an open set ${\cal O}$ of 
flow lines $\gamma = \gamma(\tau)$ of either $\nabla u$ or $-\nabla u$ 
which start on a set of positive measure on $S$ and diverge to infinity 
in $E$. Consider the former case, which corresponds to $m_{E} > 0$. A 
generic flow line of $\nabla u$ in $E$ tends to the maximal value of 
$u$ in $E$ and hence a generic flow line in ${\cal O}$ also tends this 
maximal value. By (3.42), it then follows that sup$_{E \setminus K} u = 
1$, for some compact set $K \subset E$. The function $v = - u$ is thus 
a bounded (non-constant) harmonic function on $E$, which tends 
uniformly to its infimum at infinity. Hence $E$ cannot be parabolic. 
This contradiction shows that (3.35) cannot hold for $E$, and thus, by 
the proofs in Cases A and B above, the end $E$ is asymptotically flat. 
The proof in case $m_{E} < $ 0 is the same.

  This completes the proof of Theorem 0.3.
{\endproof}

\medskip
\noindent
{\bf Remark 3.8.(i).}
  There are (non-flat) static vacuum solutions with a small end, namely 
the Kasner metrics (2.33), with $a > $ 0. These solutions have $volB(s) 
\sim  s^{2-\delta},$ $volS(s) \sim  s^{1-\delta},$ where $\delta = (a + 
a^{-1} - 1)^{-1} \in (0,1)$, and hence the end is small. Note that $u 
\sim  s^{\delta}$ is unbounded. There are other Weyl solutions which 
are complete away from $\Sigma $ with $\partial M$ pseudo-compact and 
with one small end, (take for instance the potential given by the 
Green's function on ${\Bbb R}^{2}\times S^1$, see (ii) below), but all 
known examples with small ends are either asymptotic to the Kasner 
metric at infinity or have faster than quadratic curvature decay, i.e. 
satisfy (3.18).

  It is an open problem to understand in more detail the structure of 
small ends of static vacuum solutions. It follows from the results 
above in \S 3 that all tangent cones at infinity of $E$ are collapsing, 
and hence they are all Weyl solutions. But the metric uniqueness of 
tangent cones at infinity is unknown, as is the question of whether 
small ends have finite topological type.

{\bf (ii).}
 We construct an example which illustrates the sharpness of the last 
statement of Theorem 0.3. Let $G_{1}$ and $G_{2}$ be the Green's 
functions for the Laplacian on the flat product ${\Bbb R}^{2}\times 
S^{1}$, with poles at  $(0, p_{i})$, for $p_{1}$,$p_{2}$ distinct 
points in $S^{1}$. Here we consider $S^{1}$ as the $z$-axis in ${\Bbb 
R}^{3}$ quotiented out by an isometric ${\Bbb Z}$-action. As in (2.34), 
$G_{i}(x) = G(x,p_{i})$, viewed as a function on the universal cover 
${\Bbb R}^{3}$, may be written as
\begin{equation} \label{3.46}
G_{i} = lim_{n \rightarrow \infty} \bigl[ \sum _{j=-n}^n 
\frac{1}{r_{j}} - c_{n} \bigr] ,
\end{equation}
where $r_{j}$ is the Euclidean distance to the collection of lifts of 
$p_{i}$ in ${\Bbb R}^3$ and $c_{n}$ is a suitable normalizing constant 
with $c_{n} \rightarrow \infty$ as $n \rightarrow \infty$, chosen so 
that $G(x, p_{i})$ is finite.
  Thus $G_{i}$ is an axisymmetric and $z$-periodic harmonic function on 
$({\Bbb R}^{2})^{+} \times S^{1}$, where the $S^{1}$ now means 
rotations about the $z$-axis in ${\Bbb R}^{3}$. Hence the potential
$$ \nu = G_{1} - G_{2} $$
generates a Weyl solution as in (2.12), with $u = e^{\nu}$  and which 
has an isometric ${\Bbb Z}$-action along the $z$-axis. Let $M$ denote 
the ${\Bbb Z}$ quotient of this solution. Then the metric boundary of 
$M$ is pseudo-compact. Since $G_{i} \sim log r$ as $r \rightarrow 
\infty $, 
$$u \rightarrow 1,$$ 
at infinity in $M$. The end $E$ = ${\Bbb R}^+\times T^2$ is small and 
has mass 0 in the sense of (0.7).
  These solutions of course resemble the dipole-type solutions 
discussed in \S 2(IIC), but with a collapsing end. 

\medskip
\noindent
{\bf Remark 3.9.(i).}
  Although we will not detail it here, an examination of the proof 
shows that Theorem 0.3 holds for non-vacuum static solutions of the 
Einstein field equations (1.2), provided suitable decay conditions are 
imposed on the stress-energy tensor $T$.

 This is the case for example, if $T$ is any tensor with compact 
support, or more generally if $T$ satisfies an estimate of the form 
$|T|(x) \leq  c\cdot  t^{-3}(x),$ for some $c > $ 0 and all $x$ with 
$t(x) \geq  s_{o}$, together with $\frac{1}{u}\Delta u \in  
L^{1}(M\setminus U).$ This latter condition is needed to obtain the 
bound (3.10). By (1.2), note that $\frac{1}{u}\Delta u = 
\frac{1}{2}trT.$ The starting estimate (3.3) may be obtained by 
applying (a suitable version of) [An1,Thm.3.3].

  {\bf (ii).}
    Also, Theorem 0.3 can be given a finite or quantitative 
formulation, i.e. one can relax the assumption of completeness, in 
basically the same way as the local estimates (3.3) follow from the 
non-existence of global static vacuum solutions with $\partial M = 
\emptyset$, c.f. [An1,App.].

   Thus, if (3.15) holds and $(M, g)$ is 'sufficiently large`, 
depending on $K$, then sufficiently far out in $(M, g)$, the metric is 
close to a flat metric. We leave a precise formulation and proof, 
(based on Theorem 0.3), to the reader.

\medskip
\noindent
{\bf Remark 3.10.}
  As noted in \S 0, it would be of interest to prove that $(M, g, u)$ 
has a unique end. Under the hypotheses of Theorem 0.3, we conjecture 
this is the case at least when $M$ is complete away from $\Sigma .$ If 
$M$ is in addition smooth up to $\Sigma ,$ this has been proved by 
Galloway [G]. Following essentially the same arguments as in [G], it is 
not difficult to show that if $M$ is complete away from $\Sigma $ and 
if the Riemannian 4-manifold $N$ admits a compact smoothing of a 
neighborhood of $\Sigma $ having non-negative Ricci curvature, then $M$ 
has a unique end.

\medskip
\noindent
{\bf Remark 3.11.}
  We point out that Theorem 0.3 and Theorem 0.1 are false in higher 
dimensions, due to the existence of Einstein metrics on compact 
manifolds, which are not of constant curvature, in dimensions $\geq $ 
3. (The equations (0.1) on any $n$-dimensional manifold $M^{n}$ 
generate Ricci-flat manifolds on $N^{n+1}).$ 

 Thus, let $(\Sigma , g)$ be any compact $(n- 2)$ dimensional Einstein 
manifold, with $Ric_{g} = (n- 3)\cdot  g$ and define the warped product 
metric $\Roof{g}{\bar}$ on ${\Bbb R}^{2}\times {\Sigma}$ by
$$\Roof{g}{\bar} = dt^{2} + \frac{4(f' (t))^{2}}{(n- 2)^{2}}d\phi^{2} + 
f^{2}(t)\cdot  g , $$
where $f$ is the unique function on $[0,\infty )$ such that $f(0) =$ 1, 
$f'  > $ 0 and $(f')^{2} =$ 1 $-  f^{1-n}.$ A simple computation shows 
that $({\Bbb R}^{2}\times {\Sigma} , \Roof{g}{\bar})$ is complete and 
Ricci-flat, c.f. [Bes, p.271] and the space-like hypersurface ${\Bbb 
R}^{+}\times {\Sigma} ,$ with metric $dt^{2} + f^{2}(t)\cdot  g,$ is a 
solution to the static vacuum equations, with potential $u = f' ,$ (up 
to a multiplicative constant). Thus the horizon is $\Sigma $ and the 
solution is smooth up to $\Sigma $ and complete away from $\Sigma .$

 This metric is asymptotically conical, i.e. asymptotic to the complete 
(Euclidean) cone on $(\Sigma , g)$, but is asymptotically flat only in 
the case that $(\Sigma , g) = S^{n-2}(1),$ corresponding to the 
$n$-dimensional Schwarzschild metric.

\input{ref.tex}

\input{tail.tex}

\end{document}

%% file: ref.tex
\bibliographystyle{plain}

%% file: tail.tex
\bigskip

\begin{center}
July, 1998: revision July, 2000
\end{center}
\medskip
\address{Department of Mathematics\\
S.U.N.Y. at Stony Brook\\
Stony Brook, N.Y. 11794-3651}\\
\email{anderson@@math.sunysb.edu}